\documentclass[journal]{IEEEtran}
\usepackage{siunitx}
\usepackage{amsfonts,amsmath,amsthm}
\usepackage{graphicx,xspace,epsfig,syntonly,dsfont}
\usepackage{url,cite,bm,bbm}
\usepackage{algorithmicx}
\usepackage{arydshln}
\usepackage{booktabs}
\usepackage{balance}
\usepackage{subfigure}
\usepackage{stfloats}
\usepackage{diagbox}
\usepackage{multirow}
\usepackage{enumitem}
\newcommand{\xmark}{\ding{55}}
\usepackage{pifont}
\newtheorem{remark}{Remark}

\usepackage{algorithm}
\usepackage{algpseudocode}

\long\def\symbolfootnote[#1]#2{\begingroup
\def\thefootnote{\fnsymbol{footnote}}
\footnote[#1]{#2}\endgroup}
\usepackage{xcolor}

\newcommand{\blue}[1]{{#1}}

\DeclareSIUnit \voltampere {MVA} 

\IEEEoverridecommandlockouts

\allowdisplaybreaks[4]

\title{Physics-Informed Gradient Estimation for Accelerating Deep Learning based AC-OPF}

\author{
    \IEEEauthorblockN{Kejun Chen, Shourya Bose, and Yu Zhang~\IEEEmembership{Member,~IEEE}}
    \thanks{The authors are with the Department of Electrical and Computer Engineering at the University of California, Santa Cruz, USA. Emails: \texttt\{kchen158, shbose, zhangy\}@ucsc.edu. 
    
    This work was supported in part by the UCSC Interdisciplinary Innovation Program (I2P).}. 
}

\begin{document}
\maketitle
\begin{abstract}
The optimal power flow (OPF) problem can be rapidly and reliably solved by employing responsive online solvers based on neural networks. The dynamic nature of renewable energy generation and the variability of power grid conditions necessitate frequent neural network updates with new data instances. To address this need and reduce the time required for data preparation time, we propose a semi-supervised learning framework aided by data augmentation. In this context, ridge regression replaces the traditional solver, facilitating swift prediction of optimal solutions for the given input load demands. Additionally, to accelerate the backpropagation during training, we develop novel batch-mean gradient estimation approaches along with a reduced branch set to alleviate the complexity of gradient computation. Numerical simulations demonstrate that our neural network, equipped with the proposed gradient estimators, consistently achieves feasible and near-optimal solutions. These results underline the effectiveness of our approach for practical implementation in real-time OPF applications.

\end{abstract}


\section{Introduction}\label{sect:intro}
Timely clearing of real-time energy markets is an important task carried out by independent system operators (ISOs). When the constraints of this resource allocation problem encompass nonlinear equations that represent network physics, it is referred to as alternating-current optimal power flow (AC-OPF). In practical applications, the computational burden associated with solving AC-OPF compels ISOs to resort to linear approximations, such as DC-OPF, to meet real-time operating requirements. However, the rising penetration of renewables and fluctuations in load demand make it imperative to employ AC-OPF. This is crucial to prevent network losses that may arise from the low-fidelity modeling inherent in DC-OPF. Consequently, there is a need to explore alternatives to traditional OPF solvers, as they are unsuitable for real-time scenarios \cite{Mohammadi,Hai-Tao}.

In recent times, there has been a surge in research exploring deep learning-based approaches, driven by their notable generalization capabilities and efficient inference times. Considering the data-driven nature, it is critical to frequently and promptly update the neural network (NN) with new data instances \cite{Chatzos2022}. In this paper, we propose a semi-supervised learning framework to shorten data preparation times compared to the current supervised learning literature. Additionally, we propose a novel physics-informed gradient estimation technique to accelerate backpropagation during NN training. The cumulative impact results in the swift assessment of AC-OPF, concurrently maintaining the underlying neural network updated through frequent re-training. 

\subsection{Literature Review}

Both supervised and unsupervised learning techniques have been applied to address the AC-OPF problem. In the supervised learning framework, traditional solvers are commonly used to generate input-output data pairs by finding the optimal solution to the given load demand. \cite{Fioretto2019} and \cite{chatzos2020} utilize fully connected neural networks (FCNNs) to output all decision variables. \cite{Chatzos2022} proposes a two-stage approach for scalable learning while \cite{Seonho2023} employs compact optimization learning to compress the space of optimal solutions. By contrast, \cite{Huang2021_v} and \cite{Lei2021} focus on outputting voltage phasors and recovering active/reactive power generation via power balance equations. Some studies leverage the power grid's topology information by using graph neural networks, convolutional neural networks (CNNs), and Chebyshev CNNs in place of FCNNs; see \cite{Owerko2020,falconer2022,Gao2023}. However, a major drawback of these approaches is the lack of constraint satisfaction. \cite{NELLIKKATH} proposes a physics-informed NN that outputs the primal and dual variables and adds the KKT condition violations in the loss term. To address the load mismatch issue, \cite{deepopf} and \cite{WANG2022} utilize FCNNs to predict a partial set of decision variables and reconstruct the remaining variables using power flow (PF) solvers. Furthermore, the time required for data pair preparation in existing works remains substantial due to the long solve times of conventional solvers generating the data. Frequent training on such data is important to ensure that the quality of NN solutions does not degrade~\cite{Zhou2023}. In the modern smart grid, the Internet of Things framework allows ISOs to rapidly acquire system states through the advanced communication network, such as monitoring the load demands and controlling distributed energy resources by smart devices \cite{ZhangIot}.

Recent research has used unsupervised learning as a tool to streamline the laborious process of data preparation. In unsupervised learning techniques, the training process of NNs itself ensures that its outputs are feasible and optimal solutions to AC-OPF without requiring explicit targets. \cite{huang2021} and \cite{Junfei2022} incorporate both the OPF objective function and a penalty for constraint violations in their training loss. Other approaches, such as \cite{parkaaai}, exploit duality theory to jointly train primal and dual networks with a loss function inspired by the augmented Lagrangian method. \cite{Cao} formulates the multi-period OPF as a stochastic nonlinear programming problem and solves the Markov decision process by employing reinforcement learning. To fulfill load balance, \cite{donti2021dc3} and \cite{KejunGlobal} employ a variable splitting (VS) scheme, which predicts a subset of decision variables and reconstructs the remaining variables using PF solvers (e.g., fast decoupled power flow (FDPF)). The unsupervised learning framework typically relies on bounded activation functions in the output layer to ensure the automatic satisfaction of inequality constraints. However, in the worst-case scenario, subsequent PF solvers may fail to find feasible solutions, which leads to unsuccessful training. However, variable splitting introduces new challenges when it comes to computing gradients. During each training iteration, the gradient of the reconstructed variables with respect to the predicted variables needs to be calculated. Based on the implicit function theorem, \cite{donti2021dc3} employs the inverse Jacobian matrix to derive the gradient. To alleviate the computational complexity, \cite{deepopf} and \cite{WANG2022} adopt the zeroth-order estimation method \cite{Liu2020}. Nevertheless, training can still be time-consuming, as it requires more iterations to converge to the optimal solution. Consequently, these existing schemes are ill-suited for timely and frequent updates of the NNs.

\subsection{Contributions}
To reduce the time required for data preparation and enhance the quality of the deep learning-based OPF solution, this study introduces a semi-supervised learning framework incorporating data augmentation techniques. Semi-supervised learning has been applied to detecting events and diagnosing faults in scenarios where labeled data are missing or imbalanced \cite{yangf2023} and \cite{Farajzadeh}.  In AC-OPF, iteration-based optimization solvers are typically utilized to compute the target variables. Relying on the conventional solver to generate thousands of data pairs imposes an additional heavy burden on the NN-based supervised learning framework. Thus, inspired by the pseudo-labeling technique, we construct a hybrid training dataset comprising ground truth data given by a conventional solver and pseudo data obtained through a data-driven regression. By utilizing a limited set of input-output data pairs obtained from a traditional solver, we employ ridge regression to learn the mapping from load demands to optimal decision variables. Subsequently, the trained model is used to predict pseudo target values for other demand samples. 

Moreover, during the initial training epochs, we incorporate the $\ell_2$ loss function as part of warm-up training. Following the warm-up period, the training loss is adjusted to include a penalty for constraint violations, which promotes feasibility. In addition, we address the computational and memory storage challenges associated with the Jacobian tensor
by introducing a novel batch gradient estimator. Additionally, we curtail unnecessary computations by excluding branches (power lines) that are unlikely to be binding
to their flow limits; this is done before training commences. This design is motivated by the fact that a significant number of apparent branch flow constraints are non-binding (cf. \cite{deka2019} and \cite{robson2020}). Computing gradients for these non-binding constraints is unnecessary since their violation loss is zero and does not impact weight updates.

To sum up, the major contributions of this paper are threefold: 
(1) A semi-supervised learning framework with pseudo-labeling is proposed to significantly reduce data preparation time. (2) The proposed warm-up training epochs facilitate finding feasible PF solutions even in the initial training stage, thereby expediting the neural network training process. (3) Efficient batch gradient estimation techniques, along with a reduced branch set, are developed to alleviate the high computation burden of the training. 
    
The remaining part of this paper is organized as follows. Section \ref{sec:pf} formulates the AC-OPF problem. Section \ref{sec:approach} details the proposed learning framework and approaches. Section \ref{sec:tests} shows the numerical results tested on four benchmark systems. Finally, Section \ref{sec:conclusion} presents the concluding remark.

\textit{Notation}: Upper (lower) boldface letters are used for matrices (column vectors). Sets are denoted by calligraphic letters. $(\cdot)^{\top}$ and $(\cdot)^{-1}$ are vector/matrix transpose and matrix inverse, respectively. $\|\cdot\|_2$ denotes the vector $\ell_2$-norm. 

\section{AC-OPF problem formulation} \label{sec:pf}
Consider a power network consisting of $N$ buses (denoted by $\mathcal{N}$) and $M$ power lines (denoted by $\mathcal{M}$). There are three types of buses: one reference bus, the set of $N_d$ load buses (denoted by $\mathcal{N}_d$), and the set of $N_g$ generator buses (denoted by $\mathcal{N}_g$). Let $\overline{\mathcal{N}}_d := \mathcal{N} \setminus \mathcal{N}_d$ collect all generator buses and the reference bus. 
Our AC-OPF formulation minimizes a single objective function, such as total generation cost, while satisfying a set of operational constraints, as given below:
\begin{subequations}
\label{AC-PF}
\begin{align}
    \min_{\mathbf{V}, \boldsymbol{\theta}, \mathbf{P}_g, \mathbf{Q}_g} & \sum_{i} c_i(P_{g, i}) \, , \label{oj}\\
    \textrm{s.t.} \quad P_{g, i} \!&=\! P_{d, i} \!+\! V_i \sum_{j=1}^N V_j (G_{ij}\cos \theta_{ij} \!+\! B_{ij}\sin \theta_{ij} ), \label{pf1} \\ 
    Q_{g, i} \! &= \!Q_{d, i} \!+\! V_i \sum_{j=1}^N V_j (G_{ij}\sin \theta_{ij} \!- \!B_{ij}\cos \theta_{ij} ), \label{pf2}
    \\ 
    p_{ij} \!&=\! - G_{ij} V_i^2 \!+\! V_iV_j(G_{ij}\cos \theta_{ij} \!+\! B_{ij}\sin \theta_{ij}) \, , \label{bf1}
    \\
    q_{ij} \!&=\! B_{ij} V_i^2 \!+\! V_i V_j(G_{ij}\sin \theta_{ij} \!-\! B_{ij}\cos \theta_{ij}) \, , \label{bf2} 
    \\
    s_{ij}^2 &= p_{ij}^2 \!+\! q_{ij}^2,\, \forall  (i, j)\in \mathcal{M} \, , \label{bf3}
    \\
    s_{ij}^2 &\leq (s_{ij}^{{\max}})^2,\, \forall  (i, j)\in \mathcal{M} \, , \label{bf4} 
    \\
    P_{g, i}^{\min} \!&\leq\! P_{g, i} \leq P_{g, i}^{{\max}},\, \forall  i\in \overline{\mathcal{N}}_d \, ,  \label{pg}\\
    Q_{g, i}^{\min} \!&\leq\! Q_{g, i} \leq Q_{g, i}^{\max},\, \forall  i\in \overline{\mathcal{N}}_d \, , \label{qg}\\
    V_{i}^{\min} \!&\leq\! V_{i} \leq V_{i}^{{\max}},\, \forall  i\in \mathcal{N} \, , \label{v} \\
    P_{g, i} \!&=\! Q_{g, i} = 0\, , \forall  i \in \mathcal{N}_d \, , \label{no_gen} \\
    \theta_{\text{ref}} \!&=\! 0 \, . \label{angle_ref}
\end{align} 
\end{subequations}
The objective function \eqref{oj} captures the total generation cost. $c_i(\cdot)$ is the generation cost function of generator $i$. $P_{g, i}$, $Q_{g, i}$, $P_{d, i}$ and $Q_{d, i}$ denote the active/reactive power generations and load demands of bus $i$. $V_i$ is the voltage magnitude of bus $i$. $\theta_{ij} := \theta_i - \theta_j$ is the voltage angle difference between bus $i$ and $j$. $p_{ij}$ and $q_{ij}$ refer to the active/reactive branch flows from bus $i$ to bus $j$. $G_{ij}$ and $B_{ij}$ denote the real/imaginary parts of the $(i,j)$-th element of the nodal admittance matrix $\mathbf{Y} \in \mathbb{C}^{N\times N}$, respectively. \eqref{pf1} and \eqref{pf2} refer to the PF equations based on Kichoff's law. \eqref{bf1} and \eqref{bf2} represent the branch flow equations, and \eqref{bf4} shows the apparent power flow constraints. Besides, \eqref{pg}-\eqref{v} refer to the box constraints of active/reactive power generations and voltage magnitudes. \eqref{no_gen} indicates the load bus does not connect to any generators. \eqref{angle_ref} shows the phase angle at the reference bus, which serves as the reference value for determining the phase angles of all other buses. Without fixing one bus angle, there would be infinitely many solutions differing by a uniform phase shift.
Setting the reference bus angle to zero eliminates this redundancy, leading to a unique solution for the voltage angles. The reference value is set to 0 by convention. AC-OPF problem is highly non-linear and non-convex and cannot generally be solved in polynomial time. We propose a semi-supervised learning framework for rapid AC-OPF predictions. Our approach learns an end-to-end mapping from load demands to active/reactive power generation and voltage phasors. 
 
\section{Proposed Approach}\label{sec:approach}

\subsection{FDPF-Embedded Learning Framework}
Let $\mathbf{v}=[\boldsymbol{\theta}; \mathbf{V}] \in \mathbb{R}^{2N}$ collect all voltage phasors and $\mathbf{x} = [\mathbf{P}_d; \mathbf{Q}_d] \in \mathbb{R}^{2N}$ collect all active/reactive demands. $\mathbf{P}_g$ and $P_{g,\mathrm{ref}}$ denote active power outputs of all generator buses and the reference bus, respectively. Similarly, define $\mathbf{Q}_g$ and $Q_{g,\mathrm{ref}}$ for the reactive power counterparts.
We select $\mathbf{y} = [\mathbf{P}_g; \mathbf{V}_{\overline{\mathcal{N}}_d}] \in \mathbb{R}^{2N_g + 1}$ as the output. The remaining OPF variables are collected by $\mathbf{z}_1 = [\boldsymbol{\theta}_{\mathcal{N}_g \cup \mathcal{N}_d}; \mathbf{V}_d] \in \mathbb{R}^{2N_d + N_g}$ and $\mathbf{z}_2 = [P_{g, \mathrm{ref}}; Q_{g, \mathrm{ref}}; \mathbf{Q}_g; \mathbf{s}^2] \in \mathbb{R}^{2+N_g+M}$, where $\mathbf{s}^2=[s_1^2,s_2^2,\ldots, s_M^2]^{\top}$ are the squared apparent power flows. 

\begin{figure}[tb!]
    \centering
    \includegraphics[width=0.9\linewidth]{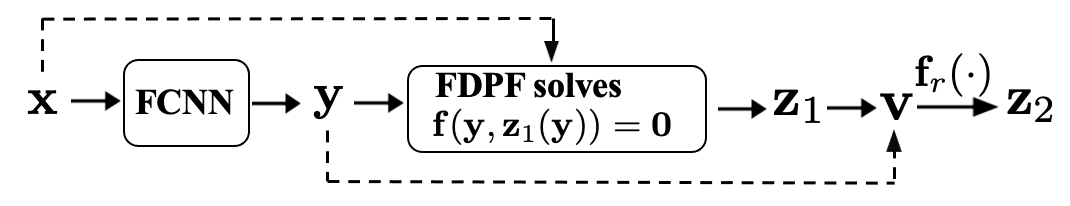}
    \caption{The proposed FCNN learning framework for the AC-OPF problem.}
    \label{opf_nn}
\end{figure}

As shown in Fig. \ref{opf_nn}, the variable splitting scheme in our proposed framework guarantees the satisfaction of the equality constraints. Specifically, given $\mathbf{x}$ and $\mathbf{y}$, the FDPF solver computes $\mathbf{z}_1$ based on the active power balance equations at load/generator buses and reactive power balance equations at load buses. They are denoted by \eqref{pf1}$_{\mathcal{N}_d \cup {\mathcal{N}_g}}$ and \eqref{pf2}$_{\mathcal{N}_d}$, respectively. We can rewrite those $2N_d + N_g$ PF equations as $\mathbf{f}(\mathbf{y}, \mathbf{z}_1(\mathbf{y})) = \mathbf{0} $. Let $ \mathbf{f}_r(\cdot)$ denote the unique mapping from $\mathbf{v}\mapsto \mathbf{z}_2$, which can be obtained by solving the equality constraints \eqref{pf1}$_{\mathrm{ref}}$, \eqref{pf2}$_{\mathrm{ref}}$, \eqref{pf1}$_{\mathcal{N}_g}$, and \eqref{bf1}-\eqref{bf3}.

We design an FCNN with $L$ layers to learn the mapping $\mathbf{x}\mapsto \mathbf{y}$. The forward propagation is given as follows:
\begin{subequations}
   \begin{align}
    \mathbf{h}_1 & = \mathbf{x} \, ,\\ 
    \mathbf{h}_{l+1} & = \sigma \left(\mathbf{W}_{l} \mathbf{h}_{l} + \mathbf{b}_{l}\right), \, l = 1, 2,\, \ldots, L-2 \, , \\
    \tilde{\mathbf{y}} &=\tanh(\mathbf{W}_{L-1}\mathbf{h}_{L-1} +  \mathbf{b}_{L-1})\, , \\
    \mathbf{y} & = \mathrm{Proj}_{\mathcal{B}_\mathbf{y}}(\tilde{\mathbf{y}}) \, ,
\end{align} 
\end{subequations}
where $\mathsf{W} := \{\mathbf{W}_l,\mathbf{b}_l\}_{l=1}^{L-1}$ collect all weights and biases. The final layer utilizes the $\tanh(\cdot)$ activation function, whereas all other layers employ the $\mathrm{ReLU}$ activation function $\sigma(\cdot)$.
Let $\mathcal{B}_{\mathbf{y}_i}:=[\mathbf{y}_i^{\min}, \mathbf{y}_i^{\max}]$ denote the interval between the minimum and maximum values of the output's $i$-th component. Operator $\mathrm{Proj}_{\mathcal{B}_\mathbf{y}}(\cdot)$ performs a component-wise projection of its argument onto $\mathcal{B}_\mathbf{y}$ given as
\begin{align}
    \mathbf{y}_i = \mathrm{Proj}_{\mathcal{B}_{\mathbf{y}_i}}(\tilde{\mathbf{y}}_i) := \lambda_i \mathbf{y}_i^{\max} + (1-\lambda_i) \mathbf{y}_i^{\min} \, ,
\end{align}
where $\lambda_i = \frac{1+\tilde{\mathbf{y}}_i}{2} \in [0,1]$ is the coefficient of the above convex combination that guarantees the fulfillment of the associated box constraint.

\subsection{Semi-supervised Learning with pseudo-labeling}
As the amount of available data increases, the accuracy of the machine learning training model improves. To speed up the training data preparation process, we combine semi-supervised learning with pseudo-labeling for deep-learning based OPF frameworks. Applying an off-the-shelf OPF solver to a small portion of demand samples, we first obtain their corresponding optimal solutions that are treated as the ground truth. Then, a regression model is trained to learn the mapping $\mathbf{x} \mapsto \mathbf{y}$. The trained model is used to predict the optimal solutions, named pseudo values, for all remaining demand samples. Those pseudo values may not be optimal or even feasible, but they play a guidance role during the training. Algorithm \ref{alg} shows the proposed process of data augmentation.

We propose to use ridge regression for the data augmentation. Given a small dataset, excessive features of the OPF problem can contribute to overfitting by introducing higher model complexity. This in turn raises the likelihood of redundant features and the inclusion of irrelevant features that have no bearing on the prediction. The primary benefit of ridge regression lies in its ability to effectively shrink coefficients and reduce model complexity. Furthermore, standard deviation (std) and ranges of OPF outputs $\mathbf{y}$ are typically small (cf. \ref{data_aug}). Ridge regression has shown promising performance in predicting outputs with small variance. 

\begin{algorithm}[tb!]
\caption{OPF data generation using pseudo-labeling}
\label{alg}
\textbf{Input:} Load demand dataset $\mathcal{X}$. \\
\textbf{Output:} Augmented load demand dataset $\hat{\mathcal{X}}$.
\begin{algorithmic}[1]
\State Use a conventional solver to find the optimal solutions to a small fraction of demand samples in $\mathcal{X}$.
\State Train a regression model to learn the input-output mapping for data points $(\mathbf{x},\mathbf{y})$ obtained from the previous step.
\State Use the trained regression model to predict pseudo values $\mathbf{y}$ for the remaining samples in $\mathcal{X}$. Project the violated pseudo values into their corresponding box constraints, i.e., $\mathbf{y} = \min \{ \max \, \{\mathbf{y}, \, \mathbf{y}^{\min}\}, \, \mathbf{y}^{\max} \}$
\State Compute $\mathbf{z}_1$ via the FDPF solver and $\mathbf{z}_2$ using equations \eqref{pf1}-\eqref{bf3}. 
\end{algorithmic}
\end{algorithm}

\subsection{Training Loss}
The training loss function should take into account the optimality and feasibility of the decision variables. Our framework shown in Fig. \ref{opf_nn} automatically guarantees the satisfaction of all equality constraints \eqref{pf1}-\eqref{bf3}. Hence, besides the generation cost $L_o$ in \eqref{oj}, we incorporate two additional terms to the training loss: i) $L_s$ as the error between pseudo values and estimates and ii) $L_c$ as the penalty of inequality constraint violation.
The overall training loss is thus designed as:
\begin{align}
\label{total_loss}
    \ell(\mathbf{y},\mathbf{z}_1,\mathbf{z}_2) := L_c + w_oL_o +  w_sL_s \, ,
\end{align}
where $w_o, w_s$ are the weighting parameters balancing different terms. The supervised learning loss $L_s$ is given as:
\begin{align}
    L_s(\mathbf{P}_g, \mathbf{v}) = \|\mathbf{P}_g - \tilde{\mathbf{P}}_g\|_2 + \| \mathbf{v} - \tilde{\mathbf{v}}\|_2 \, ,
\end{align}
where $\tilde{\mathbf{P}}_g$ and $\tilde{\mathbf{v}}$ are the pseudo values of the corresponding variables. Note that the estimation error of $\mathbf{z}_2$, which is uniquely determined by $\mathbf{v}$, is not included in $L_s$. The inequality constraint violation loss $L_c$ is constructed as: 
\begin{align}
    L_c(\mathbf{z}_2, \mathbf{V}_d) = l_c(\mathbf{z}_2) + w_vl_c(\mathbf{V}_d) \, ,
\end{align}
where $w_v$ balances the violation losses between $\mathbf{z}_2$ and $\mathbf{V}_d$. The box constraint violation loss function $l_c(\mathbf{c})$ is defined as:
\begin{align}
    l_c(\mathbf{c}) := \|\mathrm{ReLU} \, (\mathbf{c} - \mathbf{c}^{\max})\|_2 +  \|\mathrm{ReLU} \, (\mathbf{c}^{\min} - \mathbf{c})\|_2 \, . 
\end{align}

\begin{remark}[The role of weight $w_{v}$]
Using the weight parameter $w_v$ is critical in the feasibility guarantee. The motivation is to avoid the potential vicious cycle described as follows. In the backpropagation, the chain rule applies:
\begin{align}
    \frac{\mathrm{d} L_c (\mathbf{z}_2(\mathbf{V}_d), \mathbf{V}_d)}{\mathrm{d}  \mathbf{V}_d} = \frac{\partial l_c(\mathbf{z}_2)}{\partial \mathbf{z}_2} \frac{\mathrm{d} \mathbf{z}_2}{\mathrm{d} \mathbf{V}_d} + w_v\frac{\mathrm{d} \, l_c(\mathbf{V}_d) }{\mathrm{d} \mathbf{V}_d} \, ,
\end{align}
where $\frac{\mathrm{d} \mathbf{z}_2}{\mathrm{d} \mathbf{V}_d}$ is related to the power grid parameters that can have large values (see Section \ref{Jacobian model}). It is desirable to have $w_v$ and the grid parameters in a similar order of magnitude. This ensures no loss terms will be ignored in the loss function by the minimization problem because we assign similar importance to different loss terms. In the forward propagation, $\mathbf{z}_2$ depends on $\mathbf{V}_d$ through the power and branch flow equations. Thus, if $\mathbf{V}_d$ is far from the feasible region, the further obtained $\mathbf{z}_2$ will have a large violation loss. In this case, more efforts will be put into minimizing the first loss term while increasingly ignoring the second one, which leads to a vicious cycle. 
\end{remark}

\subsubsection{Warm-up epoch loss}
\label{warm-up}
In the first few warm-up epochs of training, we propose to use the loss function:
\begin{align}
\label{warm_up_loss}
    L_{\mathrm{wp}}(\mathbf{P}_g, \mathbf{V}_{\overline{\mathcal{N}}_d}) = \|\mathbf{P}_g - \tilde{\mathbf{P}}_g\|_2 + 
    w_{\mathrm{wp}}\| \mathbf{V}_{\overline{\mathcal{N}}_d} - \tilde{\mathbf{V}}_{\overline{\mathcal{N}}_d}\|_2.
\end{align}
The weight parameter $w_{\mathrm{wp}}$ balances the two loss terms. Output $\mathbf{y}$ automatically satisfies the inequality constraints thanks to the $\mathrm{Proj}$ operator. It is important to introduce the supervised training loss. First, the model is unlikely to yield good initial values of $\mathbf{y}$ during the early stage of neural network training. In the worst scenario, the FPDF solver may fail to find a feasible PF solution, leading to training failure. Second, wide feasible ranges of those decision variables lead to a longer training time. Using the neural network weights parameterized by a pre-trained model rather than random initialization helps reduce the optimality gap \cite{WANG2022}. Specifically, they pre-train an FCNN using the ground truth data to approximate the mapping $\mathbf{x}\mapsto \mathbf{y}$. We are inspired to train the neural network only with the estimation error of $\mathbf{y}$ yielding a small optimality gap. 

\subsection{Implicit Gradient Computation} 
\label{Jacobian model}
For the backpropagation, we need to calculate the derivatives $\frac{\mathrm{d} \ell}{\mathrm{d} \mathsf{W}} = \frac{\mathrm{d}\ell}{\mathrm{d}\mathbf{y}}\times \frac{\mathrm{d} \mathbf{y}}{\mathrm{d} \mathsf{W}}$, where $\frac{\mathrm{d} \mathbf{y}}{\mathrm{d} \mathsf{W}}$ is straightforward to compute. For the other factor $\frac{\mathrm{d}\ell}{\mathrm{d}\mathbf{y}}$, we have by the chain rule:
\begin{align}
    \label{chain}   \frac{\mathrm{d}\ell(\mathbf{y},\mathbf{z}_1(\mathbf{y}),\mathbf{z}_2(\mathbf{y},\mathbf{z}_1(\mathbf{y})))}{\mathrm{d}\mathbf{y}} \! = \!  \frac{\partial \ell}{\partial \mathbf{y}} \! + \! 
    \frac{\partial \ell}{\partial \mathbf{z}_1} \frac{\mathrm{d} \mathbf{z}_1}{\mathrm{d} \mathbf{y}} \! + \!  \frac{\partial \ell}{\partial \mathbf{z}_2} \! \frac{\mathrm{d} \mathbf{z}_2}{\mathrm{d} \mathbf{y}},
\end{align}
\begin{align}
    \label{chain1}
    \frac{\mathrm{d} \mathbf{z}_2(\mathbf{y},\mathbf{z}_1(\mathbf{y}))}{\mathrm{d} \mathbf{y}}  =
    \frac{\partial \mathbf{z}_2}{\partial \mathbf{y}} + \frac{\partial \mathbf{z}_2}{\partial \mathbf{z}_1}\frac{\mathrm{d}  \mathbf{z}_1}{\mathrm{d} \mathbf{y}} \, .
\end{align}
Based on the power flow and branch flow equations, $\frac{\partial \mathbf{z}_2}{\partial \mathbf{y}}$,  $\frac{\partial \mathbf{z}_2}{\partial \mathbf{z}_1}$ and $\frac{\mathrm{d} \mathbf{z}_1}{\mathrm{d} \mathbf{y}}$ can be derived from the nodal and branch Jacobian matrices as follows. 

The nodal Jacobian matrix $\mathbf{J}^{\mathrm{nodal}}$ collects the gradients of the active/reactive power injections w.r.t. the voltage phasors. It can be divided into four blocks given as:
\begin{align}
\mathbf{J}^{\mathrm{nodal}} :=
 \left[
      \begin{array}{cc}
        \mathbf{J}^{P\theta} & \mathbf{J}^{PV} \\
        \hdashline[2pt/2pt] \noalign{\vskip 0.5ex}
        \mathbf{J}^{Q\theta} & \mathbf{J}^{QV}
      \end{array} 
    \right]  \in \mathbb{R}^{2N \times 2N} \, .
    \label{eq:nodalJ}
\end{align}
Each block is an $N \times N$ matrix whose elements can be derived from the PF equations \eqref{pf1}-\eqref{pf2}.
By the apparent branch flow equation \eqref{bf3}, we have
\begin{align}
\label{branch_flow_back}
   \frac{\partial \, \mathbf{s}^2}{\partial \, \mathbf{v}_j} = 2\mathbf{p} \odot \mathbf{J}^{ab}_{(:,j)} +  2\mathbf{q} \odot \mathbf{J}^{rb}_{(:,j)} \in \mathbb{R}^{M} \, ,
\end{align}
where $\mathbf{p}=[p_1,p_2\ldots,p_M]^{\top}$ and $\mathbf{q}=[q_1,q_2\ldots,q_M]^{\top}$
are active and reactive branch flows, respectively. $\mathbf{J}^{ab} \in \mathbb{R}^{M \times 2N}$ and $\mathbf{J}^{rb} \in \mathbb{R}^{M \times 2N}$ represent the gradients of the active/reactive power flows w.r.t. the voltage phasors, which are derived by using \eqref{bf1}-\eqref{bf2}. $\mathbf{J}^{ab}_{(:,j)}$ and $\mathbf{J}^{rb}_{(:,j)}$ denote the $j$-th column of $\mathbf{J}^{ab}$ and $\mathbf{J}^{rb}$, respectively. $\mathbf{v}_j$ is the $j$-th element of $\mathbf{v}$ and $\odot$ is the point-wise product. 

Note that except for $\mathbf{P}_g$, all other variables in $\mathbf{y}$ and $\mathbf{z}_1$ are voltage phasors. The variables in $\mathbf{z}_2$ are determined by voltage phasors, we get $\frac{\partial \mathbf{z}_2}{\partial \mathbf{P}_g} =\mathbf{0}$. Furthermore, $\frac{\partial \mathbf{z}_2}{\partial \mathbf{y}}$ and $ \frac{\partial \mathbf{z}_2}{\partial \mathbf{z}_1}$ can be found as the corresponding entries of~\eqref{eq:nodalJ} an~\eqref{branch_flow_back}. 

Next, based on the implicit function theorem, we show how to compute $\frac{\mathrm{d} \mathbf{z}_1}{\mathrm{d} \mathbf{y}}$ via the PF equations $\mathbf{f}(\mathbf{y}, \mathbf{z}_1(\mathbf{y})) = \mathbf{0}$ \cite{donti2021dc3}:
\begin{align}
\label{implicit}
\frac{\mathrm{d} \mathbf{z}_1(\mathbf{y})}{\mathrm{d} \mathbf{y}} = - \left(\frac{\partial \mathbf{f}}{\partial \mathbf{z}_1} \right)^{-1} \left(\frac{\partial \mathbf{f}}{\partial \mathbf{y}} \right)  \in \mathbb{R}^{(2N_d + N_g)\times(2N_g + 1)} \, .
\end{align}
Besides, $\frac{\partial \mathbf{f}}{\partial \mathbf{z}_1}$ can be derived by extracting the corresponding elements of $\mathbf{J}^{\mathrm{nodal}}$:
\begin{align}
\frac{\partial \mathbf{f}}{\partial \mathbf{z}_1} := \mathbf{J}_{z1} = 
 \left[
      \begin{array}{cc}
        \mathbf{J}^{P\theta}_{z1} & \mathbf{J}^{PV}_{z1} \\
        \hdashline[2pt/2pt] \noalign{\vskip 0.5ex}
        \mathbf{J}^{Q\theta}_{z1} & \mathbf{J}^{QV}_{z1}
      \end{array} 
    \right] \in \mathbb{R}^{(2N_d + N_g)\times(2N_d + N_g)} \, .
\end{align}
In addition, $\frac{\partial \mathbf{f}}{\partial \mathbf{y}}$ consists of two parts: 
\begin{align}
\frac{\partial \mathbf{f}}{\partial \mathbf{y}} := \mathbf{J}_y =  
        \left[
        \begin{array}{c;{2pt/2pt}c}
        \displaystyle \frac{\partial \mathbf{f}}{\partial \mathbf{P}_g} & 
        \displaystyle  \frac{\partial \mathbf{f}}{\partial \mathbf{V}_{\overline{\mathcal{N}}_d}}
        \end{array}
        \right]  \in \mathbb{R}^{(2N_d + N_g)\times(2N_g+1)} \, .
\end{align}
For the function $\mathbf{f}$, the variables in $\mathbf{P}_g$ only get involved in \eqref{pf1}$_{\mathcal{N}_g}$. Thus, $\frac{\partial \mathbf{f}}{\partial \mathbf{P}_g} = \mathbf{C}  \in \mathbb{R}^{(2N_d + N_g)\times N_g} $ is a sparse matrix, and we have $\mathbf{C}_{ij} = -1$ when the $j$-th generator bus in $\mathbf{P}_g$ is connected to the $i$-th bus in ${\mathcal{N}_d \cup \mathcal{N}_g}$. Besides, $\frac{\partial \mathbf{f}}{\partial \mathbf{V}_{\overline{\mathcal{N}}_d}}  \in \mathbb{R}^{(2N_d + N_g)\times(N_g+1)} $ can be derived by extracting the corresponding elements of $\mathbf{J}^{\mathrm{nodal}}$.

To this end, by plugging \eqref{chain1} and \eqref{implicit} into \eqref{chain}, we get
\begin{align}
    \label{total_dev}
     \frac{\mathrm{d}\ell}{\mathrm{d}\mathbf{y}} = \frac{\partial \ell}{\partial \mathbf{y}} + \frac{\partial \ell}{\partial \mathbf{z}_2} \frac{\partial \mathbf{z}_2}{\partial \mathbf{y}} + \left(\frac{\partial \ell}{\partial \mathbf{z}_1} + \frac{\partial \ell}{\partial \mathbf{z}_2}  \frac{\partial \mathbf{z}_2}{\partial \mathbf{z}_1}\right) \left(-\mathbf{J}_{z1}^{-1}\mathbf{J}_y \right).
\end{align}
In practice, we can solve a system of linear equations instead of directly calculating the matrix inverse as follows:
\begin{align}
\label{total_dev1}
 \frac{\mathrm{d}\ell}{\mathrm{d}\mathbf{y}} = \frac{\partial \ell}{\partial \mathbf{y}} + \frac{\partial \ell}{\partial \mathbf{z}_2} \frac{\partial \mathbf{z}_2}{\partial \mathbf{y}} - \mathbf{k}^{\top} \mathbf{J}_y \, ,
\end{align}
where $\mathbf{k} \in \mathbb{R}^{2N_d + N_g}$ can be calculated by solving:
\begin{align}
\label{k}
\frac{\partial \ell}{\partial \mathbf{z}_1} + \frac{\partial \ell}{\partial \mathbf{z}_2}  \frac{\partial \mathbf{z}_2}{\partial \mathbf{z}_1} = \mathbf{k}^{\top} \mathbf{J}_{z1} \, .
\end{align}

\subsection{Proposed Batch Gradient Estimation}
During the training, the neural network employs the mini-batch gradient descent to update the weights and biases. Let $\mathcal{J}_{z1} \in \mathbb{R}^{(2N_d + N_g) \times (2N_d + N_g) \times b}$ be the tensor version of $\mathbf{J}_{z1}$, where $b$ is the batch size. Computing the batch gradient based on \eqref{total_dev1} can be practically challenging.
Given $\mathcal{J}_{z1}$, the computation complexity of solving \eqref{k} using lower–upper decomposition is $\mathcal{O}(b\times (2N_d+N_g)^3)$. It is important to note that this computation needs to be performed in every training iteration. As the scale of the power grid increases, the size of the problem grows approximately cubically, which is further amplified by the batch size. Furthermore, even with mini-batch training, memory resources can still pose a concern. While the gradient accumulation mechanism helps overcome memory limitations, it can potentially increase the training time due to the need for additional training iterations.

In this context, we introduce three models aiming to alleviate the computational burden, as elaborated below.

\subsubsection{Linearized Jacobian model} \label{Linearized Jacobian model}
The mapping from voltage phasors to the Jacobian matrix is non-linear. Linear approximations of the PF equations have been widely applied for solving PF \cite{Yang2017} and OPF problems \cite{Li2022}. We are naturally motivated to derive a linearized Jacobian formulation based on the following two assumptions:
\begin{enumerate}[label=(A\arabic*)]
    \item The phase angle difference between two connected buses is small enough. Hence, $\cos\theta_{ij} \approx 1$ and $\sin\theta_{ij} \approx \theta_{ij}$.
    \item Let ${\bar{v}}_i$ denote the mean value of the pseudo voltage magnitude at bus $i$. It is assumed that 
    \begin{align*}
    V_{i} \sin\theta_{ij} &\approx \bar{v}_{i}\theta_{ij} \, ,\\
    V_{j} \sin\theta_{ij} &\approx \bar{v}_{j}\theta_{ij} \, , \\
    V_iV_j\sin\theta_{ij} &\approx {\bar{v}}_i {\bar{v}}_j \theta_{ij} \, , \\
    V_iV_j\cos{\theta_{ij}} &\approx \bar{v}_iV_j \, . 
    \end{align*}
\end{enumerate}
Based on (A1)--(A2), the off-diagonal elements in each block of \eqref{eq:nodalJ} can be simplied as:
\begin{subequations}
\label{PF_Jac_off_lin}
\begin{align}
    J^{P\theta}_{ij} &= \bar{v}_i\bar{v}_jG_{ij} \theta_i - \bar{v}_i\bar{v}_jG_{ij} \theta_j - \bar{v}_i B_{ij} V_j \, ,  \\
    J^{PV}_{ij} &= \bar{v}_i B_{ij} \theta_{i} - \bar{v}_i B_{ij} \theta_{j} + G_{ij}V_i \, , \\
    J^{Q\theta}_{ij} &= - \bar{v}_i\bar{v}_j B_{ij} \theta_{i} + \bar{v}_i\bar{v}_j B_{ij} \theta_{j} - \bar{v}_i G_{ij} V_j \, , \\
    J^{QV}_{ij} &= \bar{v}_iG_{ij} \theta_{i} -\bar{v}_iG_{ij} \theta_{j} - B_{ij}V_i \, ,     \label{QV_ij}
\end{align} 
\end{subequations}
which can be compactly written as a linear system:
\begin{align}
\label{off_dia_items}
    \left[J^{P\theta}_{ij}\; J^{PV}_{ij}\; J^{Q\theta}_{ij}\; J^{QV}_{ij}
    \right]^{\top} 
    = \mathbf{A}_{(ij)}\mathbf{v}_{(ij)} \, ,
\end{align}
where
\begin{align}
\mathbf{v}_{(ij)} &:= \left[\theta_i\; \theta_j\; V_i\;  V_j
    \right]^{\top}, \\
    \mathbf{A}_{(ij)} &= 
    \left[
    \begin{array}{cccc}
    \bar{v}_i\bar{v}_jG_{ij} & -\bar{v}_i\bar{v}_jG_{ij} & 0 & -\bar{v}_iB_{ij} \\[3pt] 
    \bar{v}_iB_{ij} & -\bar{v}_iB_{ij} & G_{ij} & 0 \\[3pt] 
     -\bar{v}_i\bar{v}_jB_{ij} & \bar{v}_i\bar{v}_jB_{ij} &  0 & -\bar{v}_iG_{ij} \\[3pt] 
     \bar{v}_iG_{ij} & -\bar{v}_iG_{ij} & -B_{ij} & 0 
    \end{array} 
    \right] \, .
\end{align}


Similarly, all diagonal elements in the linearized Jacobian can be written as:
\begin{align}
\label{dia_items}
    \left[J^{P\theta}_{ii}\; J^{PV}_{ii}\; J^{Q\theta}_{ii}\; J^{QV}_{ii}
    \right]^{\top}
    = \mathbf{A}_{(i)} \mathbf{v} \, ,
\end{align}
where $\mathbf{A}_{(i)} \in \mathbb{R}^{4\times 2N}$ is a sparse matrix with non-zero columns given as follows. Its $i$-th and $(i+N)$-th columns are:
\begin{align}
    \mathbf{A}_{(i),\, [:,i,i+N]} = 
    \left[
    \begin{array}{cc}
    -\bar{v}_i \displaystyle \sum_{j \neq i}\bar{v}_jG_{ij}  & 0  \\[3pt] 
    \displaystyle \sum_{j \neq i}\bar{v}_jB_{ij} & 2G_{ii} \\[3pt] 
     \bar{v}_i \displaystyle \sum_{j \neq i}\bar{v}_jB_{ij}  &  0  \\[3pt] 
     \displaystyle \sum_{j \neq i}\bar{v}_jG_{ij}  & -2B_{ii}
    \end{array} 
    \right] \, ,
\end{align}
and the $j$-th and $(j+N)$-th columns (for any bus $j$ connected with bus $i$) are: 
\begin{align}
    \mathbf{A}_{(i), \, [:,j,j+N]} = 
    \left[
    \begin{array}{cc}
     \bar{v}_i\bar{v}_jG_{ij} & \bar{v}_iB_{ij} \\[3pt] 
     -\bar{v}_jB_{ij} & G_{ij} \\[3pt] 
      -\bar{v}_i\bar{v}_jB_{ij} & \bar{v}_iG_{ij} \\[3pt] 
     -\bar{v}_jG_{ij} & -B_{ij} 
    \end{array} 
    \right] \, .
\end{align}

Based on \eqref{bf1} and \eqref{bf2}, we can readily derive the linearized Jacobian for the branch flows whose details are omitted here. 
It is worth noting that all involved coefficient matrices can be pre-computed offline before the training. Thus, the computation complexity of Jacobian calculations is reduced through linearization compared to the original non-linear mappings.

\subsubsection{Decoupled Jacobian model} \label{decoupled Jacobian model}
The key idea of decoupled PF is that P-$\theta$ and Q-V have strong coupling relationships. In \eqref{k}, let $\mathbf{k}:= [\mathbf{k}_p; \mathbf{k}_q]$. We propose a decoupled formulation to calculate $\mathbf{k}_p$ and $\mathbf{k}_q$ separately by solving the following linear system:
\begin{align}
\label{k1}
\frac{\partial \ell}{\partial \mathbf{z}_1} + \frac{\partial \ell}{\partial \mathbf{z}_2}  \frac{\partial \mathbf{z}_2}{\partial \mathbf{z}_1} = \left[
      \begin{array}{c;{2pt/2pt}c}
        \mathbf{k}_p^{\top} \mathbf{J}^{P\theta}_{z1} & \mathbf{k}_q^{\top} \mathbf{J}^{QV}_{z1}
      \end{array} 
    \right] \, . 
\end{align}
The decoupled formulation \eqref{k1} solves two smaller dimensional vector $\mathbf{k}_p \in \mathbb{R}^{N_d + N_g}$ and $\mathbf{k}_q \in \mathbb{R}^{N_d}$. Hence, the computation complexity is reduced from $\mathcal{O}((2N_d+N_g)^3)$ to $\mathcal{O}((N_d+N_g)^3)$ + $\mathcal{O}(N_d^3)$.

\subsubsection{Batch-mean Jacobian tensor estimates}
During the mini-batch training, the size of Jacobian tensors $\mathcal{J}_{z1}$ and $\mathcal{J}_{v_g}$ grows quadratically with the scale of the power grid, which is amplified by the batch size. We propose a batch-mean estimation mechanism to reduce the computational burden significantly by avoiding the batch dimension. The complexity of solving \eqref{k} can be reduced from $\mathcal{O}(b\times (2N_d+N_g)^3)$ to $\mathcal{O}((2N_d+N_g)^3)$. Let $\mathcal{T} \in \mathbb{R}^{2N \times b}$ and $\mathcal{T}_{ij} \in \mathbb{R}^{4 \times b}$ represent the tensor expression of $\mathbf{v}$ and $\mathbf{v}_{(ij)}$, respectively. 
Let $\bar{\mathcal{T}}_{ij}$ and $\bar{\mathcal{T}}$ denote their mean values averaged over batch samples. Consider \eqref{off_dia_items} and \eqref{dia_items}, the estimates of batch-mean Jacobian tensors are given as
\begin{align}
    \displaystyle{\mathcal{J}_{ij}} 
    \approx \mathbf{A}_{(ij)}\bar{\mathcal{T}}_{ij}\in \mathbb{R}^{4}, \,
    \displaystyle{\mathcal{J}_{ii}}  \approx \mathbf{A}_{(i)} \bar{\mathcal{T}} \in \mathbb{R}^{4} \, .
    \label{batch_mean}
\end{align}

In the following, we will analyze the approximation error induced by the batch-mean estimation mechanism. Let $\Theta_{ij} \in \mathbb{R}^b$ and $\mathcal{V}_i \in \mathbb{R}^b$ denote the batching samples of $\theta_{ij}$ and $V_i$ whose mean values of are $\bar{\Theta}_{ij}$ and $\bar{\mathcal{V}}_{i}$, respectively. Therefore, for a given sample, the absolute errors due to the batch-mean replacement in $\mathcal{J}^{P\theta}_{ij}$ and $\mathcal{J}^{QV}_{ij}$ are given as:
\begin{align}
    e^{P\theta}_{ij} &= \left|\bar{v}_i\bar{v}_jG_{ij} (\theta_{ij} - \bar{\Theta}_{ij})
    -\bar{v}_iB_{ij} (V_j - \bar{\mathcal{V}}_j)\right| \, , \\
    e^{QV}_{ij}  &= \left|\bar{v}_iG_{ij} (\theta_{ij} - \bar{\Theta}_{ij})
    - B_{ij} (V_i - \bar{\mathcal{V}}_i)\right| \, . 
\end{align}
These errors are small due to the following facts: 1) the value of $(\theta_{ij} - \bar{\Theta}_{ij})$ is generally small; 2) the conductance value $G_{ij}$ is typically smaller than susceptance value $B_{ij}$; and 3) small value of $(V_j - \bar{\mathcal{V}}_j)$ leads to the small value of $B_{ij} (V_j - \bar{\mathcal{V}}_j)$.

In addition, the absolute error resulting from the batch-mean replacement in $\mathcal{J}^{P\theta}_{ii}$ is shown in Eq.~\eqref{error_p} and its bound is given in Eq.~\eqref{error_pb}:
\begin{align}
\label{error_p}
    e^{P\theta}_{ii} &= |\sum_{j \neq i} \bar{v}_i\bar{v}_jG_{ij} (\theta_{ij} - \bar{\Theta}_{ij})
    -\bar{v}_iB_{ij} (V_j - \bar{\mathcal{V}}_j)| \, ,\\
    e^{P\theta}_{ii} & \leq \sum_{j \neq i} e^{P\theta}_{ij}\, .\label{error_pb}
\end{align}
Similarly, the absolute error for $\mathcal{J}^{QV}_{ii}$ is given in Eq.~\eqref{error_q}:
\begin{align}
\label{error_q}
    e^{QV}_{ii} \! \leq \! |2B_{ii} (V_i \!-\! \bar{\mathcal{V}}_i)| \!+\! \sum_{j \neq i} |\bar{v}_j G_{ij} (\theta_{ij} \! - \! \bar{\Theta}_{ij}) \!-\! B_{ij}(V_j \! - \! \bar{\mathcal{V}}_j) | \, .
\end{align}
Compared with the off-diagonal entries, diagonal entries are likely to have larger error values due to the summation over $N-1$ buses. However, the summation only contains a limited number of non-zero values because the topology of a power grid is typically sparse. To this end, we have demonstrated that the proposed batch-mean estimation mechanism efficiently bypasses the tensor dimension, resulting in negligible approximation errors. 

\subsection{Reduced Branch Set}
According to \eqref{bf1}--\eqref{bf2}, the values of active/reactive branch flows can vary significantly across different batch samples of voltage phasors. Therefore, the aforementioned batch-mean replacement may not be applicable to the nonlinear mapping $\{\mathbf{p}, \mathbf{q}\} \mapsto \mathbf{s}^2$ in \eqref{bf3} (cf. \eqref{branch_flow_back} for its gradient). 
To alleviate the computational burden of branch gradients, we now propose a reduced branch set as follows. For any branch $(i, j) \in \mathcal{M}$, let $\mathcal{\tilde{S}}_{ij}^2$ collect the pseudo values of the apparent power squared. Note that as the pseudo value approaches the upper bound, the likelihood of violating the constraint increases. Thus, we define an indicator function
\begin{align}
    l_p(i, j) := \sum_{\tilde{s}_{ij} \in \mathcal{\tilde{S}}_{ij}^2}\mathrm{ReLU} \, \left( \tilde{s}_{ij}^2 - \beta (s_{ij}^{{\max}})^2 \right), \,
    \label{indi}
\end{align}
where parameter $\beta \in [0,1]$ helps quantify the likelihood of constraint violation. The indicator function is calculated for all branches in $\mathcal{M}$.

To this end, we can define the reduced branch set $\mathcal{M}_r:= \left\{(i,j)\in \mathcal{M}\, | \, l_p(i, j) \geq 0 \right\}$, which contains only $M_r = |\mathcal{M}_r|$ lines that are likely to violate the flow limit constraints. To reduce the computation burden during the training, we only calculate the gradients for those branches in $\mathcal{M}_r$. 
It is worth noting that the branches in $\mathcal{M} \setminus \mathcal{M}_r$ have a low risk of constraint violation. This is due to the inclusion of $L_s$ in the training loss, which encourages estimates to be in close proximity to the pseudo values. The summary of the proposed framework is shown in Algorithm~\ref{alg2}. 

\begin{algorithm}[tb!]
\caption{\blue{Semi-supervised OPF learning framework utilizing batch-mean gradient estimation}}
\label{alg2}
\textbf{Input:} Load demand dataset $\mathcal{X}$. \\
\textbf{Output:} The trained FCNN. 
\begin{algorithmic}[1]
\State \parbox[t]{\dimexpr\linewidth-\algorithmicindent-0em}{Construct the hybrid load demand dataset $\hat{\mathcal{X}}$ using the Algorithm~\ref{alg}. \algorithmiccomment{\texttt{Pseudo-labeling}}}
\vskip 4pt 
\State \parbox[t]{\dimexpr\linewidth-\algorithmicindent-0em}{Calculate the indicator function Eq.\eqref{indi} to identify the reduced branch set $\mathcal{M}_r$.}
\vskip 4pt 
\For {episode $e = 1,2, \dots, n_{\mathrm{tol}}$}

\Statex \parbox[t]{\dimexpr\linewidth-\algorithmicindent-1em}{\Comment{\texttt{The variable splitting scheme}}}

\State \parbox[t]{\dimexpr\linewidth-\algorithmicindent-2em}{Feed demand samples into the FCNN and retrieve the decision variables $\mathbf{y}$, $\mathbf{z}_1$, and $\mathbf{z}_2$; see Fig.~\ref{opf_nn}.}
\vskip 4pt 

\Statex \parbox[t]{\dimexpr\linewidth-\algorithmicindent-5em}{\Comment{\texttt{The branch set}}}

\If {the method $M_0$ is adopted}
\State \parbox[t]{\dimexpr\linewidth-\algorithmicindent-2em}{Compute the Jacobian of all branches in $\mathcal{M}$ using Eq.~\eqref{branch_flow_back}.}
\vskip 4pt 
\ElsIf {for the method $M_1$-$M_4$}
\State \parbox[t]{\dimexpr\linewidth-\algorithmicindent-2em}{Compute the Jacobian of branches in the reduced branch set $\mathcal{M}_r$ using Eq.~\eqref{branch_flow_back}.} \vskip 2pt 
\EndIf

\Statex \parbox[t]{\dimexpr\linewidth-\algorithmicindent-3em}{\Comment{\texttt{The Jacobian computation}}}
\vskip 4pt 
\If {for the method $M_0$, $M_1$, or $M_3$}
\State \parbox[t]{\dimexpr\linewidth-\algorithmicindent-2em}{Compute the nodal Jacobian matrix \eqref{eq:nodalJ} based on the PF equations \eqref{pf1}-\eqref{pf2}.}
\vskip 2pt 
\ElsIf {for the method $M_2$ or $M_4$}
\State \parbox[t]{\dimexpr\linewidth-\algorithmicindent-2em}{Compute the linearized Jacobian matrix using Eqs.~\eqref{off_dia_items} and~\eqref{dia_items}.}
\vskip 4pt 
\EndIf
\State \parbox[t]{\dimexpr\linewidth-\algorithmicindent-2em}{Compute batch-mean gradient using the Eq.~\eqref{batch_mean}}
\vskip 4pt 

\Statex \parbox[t]{\dimexpr\linewidth-\algorithmicindent-5em}{\Comment{\texttt{The training loss}}}
\If {$e \leq n_{\mathrm{wp}}$}: 
\State \parbox[t]{\dimexpr\linewidth-\algorithmicindent-3em}{Calculate the training loss function Eq.~\eqref{warm_up_loss} for the warm-up epochs. Jump to Line 26.}
\vskip 4pt 
\Else 
\State Calculate the training loss function Eq.~\eqref{total_loss}.
\EndIf

\Statex \parbox[t]{\dimexpr\linewidth-\algorithmicindent-2em}{\Comment{\texttt{The decoupled formulation}}}
\If {for the method $M_0$, $M_1$, or $M_2$}
\State Calculate the derivative $\frac{\mathrm{d} \ell}{\mathrm{d} \mathbf{y}}$ using Eq.~\eqref{total_dev1}. 
\ElsIf {for the method $M_3$ or $M_4$}
\State \parbox[t]{\dimexpr\linewidth-\algorithmicindent-3em}{Calculate the derivative $\frac{\mathrm{d} \ell}{\mathrm{d} \mathbf{y}}$ using the decoupled formulation Eq.~\eqref{k1}.} 
\EndIf 
\vskip 4pt 
\State \parbox[t]{\dimexpr\linewidth-\algorithmicindent-1em}{Update the weights of the FCNN through backpropagation until the training loss converges.}
\vskip 4pt 
\EndFor

\end{algorithmic}
\end{algorithm}

\section{Numerical Results} \label{sec:tests}
We evaluate the effectiveness of the proposed work on four different benchmark systems given in \texttt{MATPOWER 7.0}: IEEE-118, PEGASE-1354, PEGASE-2869, and PEGASE-9241 bus systems. The load demand data are sampled uniformly in the range of 0.8 to 1.2 times the nominal values.
\subsection{Simulation Setup}
We generate 4,000 demand samples for the PEGASE-9241 bus system and 10,000 samples for each of the other three systems. The distribution of training, validation, and test samples follows a ratio of 7:1:2. The conventional solver \texttt{MIPS} calculates the optimal decision variables for only 100 samples in the training set $\mathcal{X}$. We conduct the experiments on an iMac equipped with a 3.2 GHz CPU and 32 GB RAM. The Adam optimizer is used with \texttt{Pytorch 1.12.1}. The mini-batch size is 32. We stop training when the loss has stabilized without noticeable improvement \cite{Prechelt2012}. 

\subsection{Competing Methods}
Table \ref{Methods} summarizes state-of-the-art methods and our proposed semi-supervised learning framework. \blue{The optimal decision variables are obtained by the conventional optimization solver in the supervised learning framework. The training loss function includes the $\ell_2$ norm loss of the NN output and the labels. In the unsupervised learning framework, the output labels are not necessary. The optimality and feasibility of the decision variables are ensured by properly designing the training loss function, i.e., minimizing the generation cost and penalizing constraint violations. In addition, three VS schemes are detailed as follows:}
\begin{itemize}
    \item \blue{$\mathrm{VS}_1$ (cf. Fig.\ref{opf_nn}): predict $\mathbf{y}$ and reconstruct $\mathbf{z}_1$ using the FDPF solver. After obtaining $\mathbf{v}$, compute the remaining variables via $\mathbf{z}_2 = \mathbf{f}_r(\mathbf{v})$. The gradient of $\mathbf{z}_1$ w.r.t. $\mathbf{y}$ is implicit due to the inverse PF equations, see Eq.\eqref{total_dev}, which is the main focus of this paper. The gradient of $\mathbf{z}_2$ w.r.t. $\mathbf{v}$ is straightforward due to the explicit forward power flow and branch flow equations, i.e., a submatrix of the nodal Jacobian $\mathbf{J}^{\mathrm{nodal}}$ and the Jacobian matrix of the branch flow $\mathbf{J}^{ab}$ and $\mathbf{J}^{rb}$.} 
    
    \item \blue{$\mathrm{VS}_2$ ((cf. Fig.\ref{opf_nn2}): predict $\mathbf{v}$ and compute the remaining variables $\mathbf{P}_g$ and $\mathbf{z}_2$ based on power and branch flow equations. The gradient computation is straightforward because the mapping $\mathbf{v} \mapsto \{\mathbf{P}_g, \mathbf{z}_2$\} are explicitly described in forward power flow and branch flow equations, i.e., the nodal Jacobian matrix $\mathbf{J}^{\mathrm{nodal}}$ and the Jacobian matrix of the branch flow $\mathbf{J}^{ab}$ and $\mathbf{J}^{rb}$.}

    \begin{figure}[ht!]
    \centering
    \includegraphics[width=0.9\linewidth]{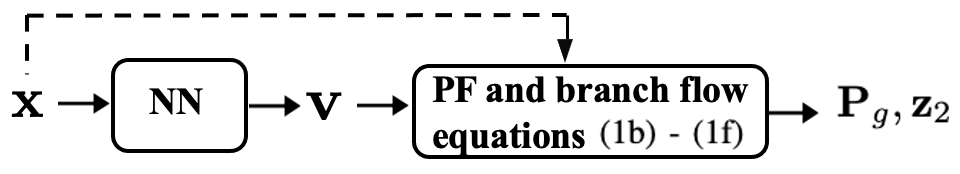}
    \caption{\blue{The NN learning framework based on $\mathrm{VS}_2$ for the AC-OPF problem.}}
    \label{opf_nn2}
    \end{figure}

    \item \blue{$\mathrm{VS}_3$: output $\mathbf{v}$, $\mathbf{P}_g$ and $\mathbf{Q}_g$ simultaneous by the NN. It does not require gradient computation among the decision variables.} 
    
\end{itemize}
Unlike $\mathrm{VS}_3$, $\mathrm{VS}_1$ and $\mathrm{VS}_2$ only predict partial decision variables and reconstruct the remaining decision variables. Besides, $\mathrm{VS}_1$ solves the inverse PF equations using the PF solver and computes the gradient implicitly and complicated. $\mathrm{VS}_2$ calculates the forward PF equations, and the gradient computation is explicit and straightforward. Note that our proposed work relies on $\mathrm{VS}_1$ because $\mathrm{VS}_2$ and $\mathrm{VS}_3$ often yield load mismatch. \blue{The proposed gradient estimation methods aim to relieve the computational burden arising from $\mathrm{VS}_1$.}

In addition, the NN weights of \emph{$M_{\mathrm{STRT}}$} are initialized from a pre-trained FCNN model, while \emph{$M_{\mathrm{FCNN}}$} initializes the weights with random values. The unsupervised learning framework \emph{${M}_{\mathrm{DC3}}$} and \emph{${M}_{\mathrm{DUAL}}$} do not require the conventional solver to generate the data pairs and thus the data preparation time is negligible. \emph{${M}_{\mathrm{DC3}}$} employs the $\ell_2$ norm as the loss function, while \emph{${M}_{\mathrm{DUAL}}$} uses the augmented Lagrangian. \emph{${M}_{\mathrm{CHC}}$} and \emph{${M}_{\mathrm{CNN}}$} use CNNs to utilize the topology information of the power grid. In addition, \emph{${M}_{\mathrm{COMP}}$} incorporates principal component analysis to compress the FCNN output dimension. Besides, they use post-processing to obtain the remaining decision variables $\mathbf{z}_1$ and $\mathbf{z}_2$ upon the completion of training. By contrast, \emph{$M_{\mathrm{FCNN}}$} reconstructs $\mathbf{z}_1$ and $\mathbf{z}_2$ to calculate the loss function during training. These variables require the associated gradient computation. Additionally, $M_1$ is compared with $M_0$ to show the advantage of the reduced branch set. The effectiveness of the linearized and decoupled models is verified by comparing $M_2$ and $M_3$ with $M_1$, respectively. By combining the linearized and decoupled Jacobian models, $M_4$ has the least computational burden.

\begin{table*}
\caption{\blue{Methods for comparison. The second column shows VS schemes. The third column shows NN structures. The last column summarizes the gradient computation methods and the gradient estimation methods.}}
\label{Methods}
\centering
\scalebox{1.1}{
{\begin{tabular}{l|l|l|l|l}\hline 
Framework & VS & NN & Method & Gradient computation method\\
\cline{1-5}
\multirow{8}{*}{Supervised} & \multirow{4}{*}{$\mathrm{VS}_1$} &  \multirow{2}{*}{FCNN} & \emph{$M_{\mathrm{FCNN}}$} \cite{deepopf} &  Zeroth-order estimation based on the training loss function \\  
\cline{4-5}  
& & & \emph{${M}_{\mathrm{STRT}}$} \cite{WANG2022} & Zeroth-order estimation based on the training loss function  \\
\cline{3-5}  
& & Chebyshev& \emph{${M}_{\mathrm{CHC}}$} \cite{falconer2022} & Post-processing with the PF solver, no extra gradient computation required \\
\cline{3-5} 
& &  CNN & \emph{${M}_{\mathrm{CNN}}$} \cite{falconer2022} & Post-processing with the PF solver, no extra gradient computation required \\
\cline{2-5}

& \multirow{1}{*}{$\mathrm{VS}_2$} & FCNN & \emph{${M}_{\mathrm{FCNNV}}$} \cite{Huang2021_v} & Explicit nodal Jacobian matrix using the forward PF equations; see Eq.\eqref{eq:nodalJ}\\  
\cline{2-5}
& \multirow{2}{*}{$\mathrm{VS}_3$} & FCNN & \emph{${M}_{\mathrm{COMP}}$} \cite{Seonho2023} & Post-processing with the PF solver, no extra gradient computation required \\ 
\cline{3-5}
& & LSTM & \emph{${M}_{\mathrm{LSTM}}$} & Post-processing with the PF solver, no extra gradient computation required \\
\hline

\multirow{2}{*}{Unsupervised} 
& \multirow{2}{*}{$\mathrm{VS}_1$} 
& \multirow{2}{*}{FCNN}  & \emph{${M}_{\mathrm{DC3}}$} \cite{donti2021dc3} & Implicit Jacobian based on the inverse PF equations; see Eq.\eqref{eq:nodalJ} and Eq.\eqref{total_dev}  \\
\cline{4-5}
& &  & \emph{${M}_{\mathrm{DUAL}}$} \cite{KejunGlobal} & Implicit Jacobian based on the inverse PF equations; see Eq.\eqref{eq:nodalJ} and Eq.\eqref{total_dev} \\
\cline{2-5}
& $\mathrm{VS}_2$ & FCNN & \emph{${M}_{\mathrm{NGT}}$} \cite{huang2021} & Explicit nodal Jacobian matrix using the forward PF equations; see Eq.\eqref{eq:nodalJ} \\

\cline{1-5}

\multirow{5}{*}{Semi-supervised} & \multirow{5}{*}{$\mathrm{VS}_1$} & \multirow{5}{*}{FCNN} & $M_0$ & Batch-mean estimation with all branches \\
\cline{4-5}
& & & $M_1$ & Batch-mean estimation \& reduced branch set \\ 
\cline{4-5}
& & & $M_2$ & Linearized Jacobian \& batch-mean estimation \& reduced branch set\\ 
\cline{4-5}
& & &  $M_3$ & Decoupled Jacobian \& batch-mean estimation \& reduced branch set\\ 
\cline{4-5}
& & & $M_4$ & Linearized decoupled Jacobian \& batch-mean estimation \& reduced branch set \\ 
\cline{1-5}
\end{tabular}}
}
\end{table*}

\subsection{Performance Criteria}
The performance of the proposed approaches is evaluated by using the following metrics.
\begin{itemize}
    \item \textbf{Optimality gap:} let $C_{\mathrm{}}$ and $C_o$ denote the optimal cost given by the NN and the solver \texttt{MIPS}, respectively. The optimality gap is $
        l_{\mathrm{cost}} = \frac{C_{\mathrm{}} -C_o}{C_o} \times 100\%$.  
    \item \textbf{Feasibility:} let $\mathbf{z}_a = [\mathbf{P}_g; P_{g,\mathrm{ref}}; \mathbf{Q}_g; Q_{g,\mathrm{ref}}; \mathbf{V}; \mathbf{s}^2]$ collect quantities for all branches. The inequality constraint violation is given by $ l_v(\mathbf{z}_a ) = \mathrm{ReLU} \left(\mathbf{z}_a - \mathbf{z}_a^{\max}\right) + \mathrm{ReLU} \left(\mathbf{z}_a^{\min} - \mathbf{z}_a \right)$. Its maximum and mean values are denoted as $l_v^{\max}$ and $\bar{l}_v$, respectively.   
    \item \textbf{Load mismatch:} ratio of the absolute error of load demand estimate to the ground truth, denoted by $e_l$.    
    \item \textbf{Computational time:} let $t_\mathrm{train}$ and $T_\mathrm{train}$ represent the training time of each epoch and the total training time. $T_\mathrm{prop}$ and $T_\mathrm{opt}$ denote the testing time of our proposed methods and the conventional optimization solver, respectively.
    \item \textbf{Storage:} required memory size of the data.    
\end{itemize}

\subsection{Simulation Results}
\subsubsection{Data preparation}\label{data_aug}
\begin{table}[tb!]
\caption{The average std values and the ridge regression parameters of the decision variables.}
\label{Stage1_std}
\centering
\begin{tabular}{c|c|c|c|c}
\hline 
System & std ($\mathbf{P}_g$) & std ($ \mathbf{V}_{\overline{\mathcal{N}}_d}$) & $\alpha_p$ & $\alpha_v$\\ \hline
118 & 0.032 & 0.002 & 0.01 & 0.1\\
1354 & 0.067 & 0.001 & 1 & 1 \\
2869 & 0.082 & 0.001 & 10 & 10 \\
9241 & 0.062 & 0.003 & 100 & 500 \\
\hline
\end{tabular}
\end{table}
Given 100 pairs of the ground truth data, we split them into the training and validation dataset with a ratio of 8:2. As shown in Table \ref{Stage1_std}, the average std values of $\mathbf{P}_g$ and $\mathbf{V}_{\overline{\mathcal{N}}_d}$ are small. For ridge regression, $\alpha_p$ and $\alpha_v$ represent the hyper-parameters that control the regularization strength of $\mathbf{P}_g$ and $\mathbf{V}_{\overline{\mathcal{N}}_d}$, respectively. In addition, Table \ref{Stage1_error} shows the validation errors between ridge regression estimates and ground truth are small. Thus, it is reliable to construct the pre-computed coefficient matrix based on the pseudo values and incorporate the supervised training loss $L_s$. Finally, Table \ref{Methods_time} shows that the proposed semi-supervised learning framework significantly reduces the data preparation time compared with the supervised learning counterpart. 

\begin{table}[tb!]
\caption{The $\ell_2$ norm of prediction errors of the ridge regression.}
\label{Stage1_error}
\centering
\begin{tabular}{c|c|c|c|c}
\hline 
System & $\ell_2(\mathbf{P}_g)$ & $\ell_2(\mathbf{Q}_g)$  & $\ell_2(\mathbf{V})$ & $\ell_2(\boldsymbol{\theta})$  \\ \hline
118 & 0.165 & 0.278 & 0.018 & 0.219  \\

1354 & 1.828 & 1.235 & 0.042 & 0.691 \\

2869 & 4.279 & 2.061 & 0.083 & 2.993 \\

9241 & 5.599 & 4.615 & 0.406 & 8.825 \\
\hline
\end{tabular}
\end{table}

\begin{table}[tb!]
\caption{Training data preparation time.}
\label{Methods_time}
\centering
\begin{tabular}{c|l|l}\hline 
System & Learning framework & Time  \\
\hline
\multirow{3}{*}{118} & Supervised & 0.2h \\  
& Unsupervised & 0.02s\\
& Semi-supervised & 0.2min \\  
\hline 
\multirow{3}{*}{1354} & Supervised & 2.2h \\  
& Unsupervised  & 0.3s \\
& Semi-supervised & 3min \\  
\hline 

\multirow{3}{*}{2869} & Supervised  & 6.4h \\   
& Unsupervised  & 0.6s \\ 
& Semi-supervised  & 9min \\  
\hline 

\multirow{3}{*}{9241} & Supervised  & 12.2h \\   
& Unsupervised  & 0.9s \\ 
& Semi-supervised  & 20.5min\\  
\hline 

\end{tabular}
\end{table}

\subsubsection{Training setup}
Table \ref{Stage1_error} shows the orders of magnitude of the different loss values, which can guide us to tune the weight parameters in the joint training loss function. Besides, as shown in \eqref{QV_ij}, the value of $w_v$ should have a similar order of susceptance values. 
As shown in Table \ref{Stage1_reduced}, a larger value of $\beta$ leads to a smaller set $\mathcal{M}_r$. We set $\beta = 0.7 $ in the simulations. Branches in $M \setminus M_r$ do not violate the constraints, but the computational burden of the training is greatly reduced. In addition, the number of warm-up epochs is determined when no significant change of $L_{\mathrm{wp}}$ can be observed during the training. The warm-up epochs play a critical role in guaranteeing that a feasible PF solution can be found by the FDPF solver in the initial training stage. Besides, we use Pytorch's \texttt{MultiStepLR} function that decays the learning rate $l_r$ by $\gamma$ once the number of epochs reaches the milestone. Given limited training time, we adopt the learning rate schedule strategy to improve training convergence and stability. A large learning rate is employed at the beginning of training to accelerate the learning process. Towards the later stages of training, we reduce the learning rate to prevent oscillations and facilitate convergence to a local minimum. Finally, Table~\ref{NN_str_weight} lists the FCNN structure, the weight parameters, the number of warm-up epochs $n_\mathrm{wp}$ and total training epochs $n_\mathrm{tol}$, and the learning rate. 

\begin{table}[tb!]
\caption{The size of reduced branch set with different $\beta$. The third column shows the ratio of $M_r$ to $M$.}
\label{Stage1_reduced}
\centering
\begin{tabular}{c|c|c|c}
\hline 
System & $\beta$ & $M_r$ & $M_r/M$  \\ \hline
\multirow{5}{*}{118} & 0.9 & 29 & 0.15 \\
& 0.7 & 33 & 0.17 \\
& 0.5 & 55 & 0.29 \\
& 0.3 & 94 & 0.50 \\
\cline{1-4}

\multirow{4}{*}{1354} & 0.9 & 18 & 0.009 \\
 & 0.7 & 42 & 0.02 \\
& 0.5 & 113 & 0.05 \\
& 0.3 & 245 & 0.12  \\
\cline{1-4}

\multirow{4}{*}{2869} & 0.9 & 30 & 0.006 \\
&  0.7 & 51 & 0.01 \\
&   0.5 & 101 & 0.02 \\
&  0.3 & 243 & 0.05  \\
\cline{1-4}

\multirow{4}{*}{9241} & 0.9 & 37 & 0.002 \\
&  0.7 & 73 & 0.004 \\
&  0.5 & 134 & 0.008 \\
&  0.3 & 356 & 0.022  \\
\cline{1-4}

\end{tabular}
\end{table}

\begin{table*}[tb!]
\caption{The FCNN structure, weight parameters, the training epochs, and the learning rate of our proposed schemes.}
\label{NN_str_weight}
\centering
\begin{tabular}{c|c|c|c|c|c|c|c|c}\hline 
System & FCNN structure & $w_{\mathrm{wp}}$ & $w_v$ & $w_o$ & $w_s$ & $n_\mathrm{wp}$ & $n_\mathrm{tol}$ & Learning rate schedule ($l_r$ \& milestone \& $\gamma$)\\\hline 
118 & [236, 50, 236] & 10 & 10 & 0.1 & 0.1 & 1 & 100 & 0.0005 \& 90 \& 0.2 \\
1354 & [2708, 50, 50, 2708] & 10 & 100 & 0.01 & 0.01 & 1 & 100 & 0.0005 \& 70 \& 0.2 \\
2869 & [5738, 50, 50, 50, 5738] & 10 & 100 & 0.001 & 0.01 & 1 & 10 & 0.0005 \& 1 \& 0.1 \\
9241 & [18482, 50, 50, 50, 18482] & 10 & 100 & 0.0001 & 0.01 & 5 & 15 & 0.001 \& 5 \& 0.01 \\\hline
\end{tabular}
\end{table*}

\begin{table*}[tb!]
\caption{Comparison Results: Feasibility and Optimality. Training failure (\xmark) is indicated when the FDPF solver cannot find feasible PF solutions. The winners of methods without load mismatch are highlighted in bold.}
\label{Performance_comp}
\centering
\scalebox{0.78}{
\begin{tabular}{c|c|c|c|c|c|c|c|>{}c|c|>{}c|c||c|c|c|c|c}\hline 
System & Evaluation metrics & \emph{$M_{\mathrm{FCNN}}$} & \emph{${M}_{\mathrm{STRT}}$} & \emph{${M}_{\mathrm{FCNNV}}$} & \emph{${M}_{\mathrm{CHC}}$} & \emph{${M}_{\mathrm{CNN}}$} 
& \emph{${M}_{\mathrm{COMP}}$} & \emph{${M}_{\mathrm{LSTM}}$} & \emph{${M}_{\mathrm{DC3}}$} & \emph{${M}_{\mathrm{DUAL}}$} &  \emph{${M}_{\mathrm{NGT}}$} & $M_0$ & $M_1$ & $M_2$ & $M_3$ & $M_4$ \\ 
 \hline
\multirow{4}{*}{118} & $l_v^{\max}(10^{-1})$ & 0.33 & \textbf{0.01} & 0.89 & 0.46 & 3.05 & 2.25 & 0.53 & \textbf{0.01} & 0.05 & 0.00 & 0.04 & 0.02 & 0.03 & 0.10 & 0.06 \\

& $\bar{l}_v$($10^{-4}$) & 0.68 & \textbf{0.02} & 5.45 & 1.60 & 7.84 & 11.2 & 2.73 & \textbf{0.02} & 0.13 & 0.00 & 0.07 & 0.04 & 0.05 & 0.18 & 0.12 \\

& $l_{\mathrm{cost}}$ & 0.88 & 0.30 & 0.00 & 0.00 & 0.00 & 0.04 & 0.02 & 0.27  & 0.04 & -1.22 & 0.32 & 0.25 & 0.26 & 0.49 & 0.54 \\

& $e_l$ ($\%$) & 0 & 0 & 10 & 0 & 0 & 0 & 0 & 0 & 0 & 10 & 0 & 0 & 0 & 0 & 0  \\

\hline

\multirow{4}{*}{1354} & $l_v^{\max}(10^{-1})$ & 5.8 & 3.8 & 26.4 & 10.1 & 31.1 & 42.4 & 16.3 & 4.8 & 3.7 & 4.0 & 1.6 & \textbf{1.3} & \textbf{1.3} & \textbf{1.3} & 1.4 
\\

& $\bar{l}_v$($10^{-4}$) & 8.6 & 2.8 & 26.2 & 5.1 & 14.7 & 38.0 & 18.2 & 2.3 & 3.0 & 3.4 & 0.63 & 0.44 & \textbf{0.41} & 0.46 & 0.44 
\\

& $l_{\mathrm{cost}}$ & 0.98 & 0.03 & 0.08 & 0.00 & 0.00 & 0.00 & 0.00 & 0.76 & 0.80 & -2.99 & 0.05 & 0.05 & 0.05 & 0.06 & 0.06
\\

& $e_l$ ($\%$) & 0 & 0 & 12 & 0 & 0 & 0 & 0 & 0 & 0 & 22 & 0 & 0 & 0 & 0 & 0 \\
\hline

\multirow{4}{*}{2869} & $l_v^{\max}(10^{-1})$ & 13.5 & 6.6 & 72.3 & 32.6 & 66.4 & 72.5 & 134.6 &  6.1 & 7.0 & 3.6 & 4.1 & \textbf{3.6} & \textbf{3.6} & 4.1 & 3.9 \\

& $\bar{l}_v$($10^{-4}$) & 43.4 & 6.7 & 22.0 & 9.9 & 13.2 & 19.8 & 78.2 & 1.8 & 5.0 & 1.5 & 1.3 & \textbf{0.9} & 1.1 & 1.1 & 1.2
\\

& $l_{\mathrm{cost}}$ & 0.91 & 0.08 & 0.12 & 0.00 & 0.00 & 0.00 & 0.01 & 0.80 & 1.49 & -2.24 & 0.04 & 0.04 & 0.04 & 0.09 & 0.09
\\

& $e_l$ ($\%$) & 0 & 0 & 544 & 0 & 0 & 0 & 0 &  0 & 0 & 284 & 0 & 0 & 0 & 0 & 0   \\
\hline

\multirow{4}{*}{9241} & $l_v^{\max}(10^{-1})$ & \multirow{4}{*}{\xmark} & 165.2  &  52.6 & 18095 & \multirow{4}{*}{-} & 129.2 & 567.7 & \multirow{4}{*}{\xmark} &   \multirow{4}{*}{\xmark} & 2.2 
& 33.6 & \textbf{23.6} & 27.7 & 29.2 & 28.4 
\\  
& $\bar{l}_v$($10^{-4}$) & & 70.0 & 10.2 & 1455 & & 41.8 & 74.9 & & & 4.6 & 5.4 & 4.0 & \textbf{3.9} & 4.8 & 4.0
\\
& $l_{\mathrm{cost}}$ & & 0.01 & 0.05 & 0.03 & & 0.12 & 0.00 & & & -3.13 & 0.03 & 0.03 & 0.03 & 0.03 & 0.03
\\
& $e_l$ ($\%$) & & 0 & 484 & 0 & & 0 & 0 & & & 128 & 0 & 0 & 0 & 0 & 0
\\ 
\hline

\end{tabular}
}
\end{table*}

\begin{table}[tb!]
\caption{The training time and the memory size of the gradient data.}
\label{time}
\centering
\scalebox{0.87}{
\begin{tabular}{c|c|c|c|c|c}\hline 
System & Gradient calculation & Method & Storage & $t_\mathrm{train}$ & $T_\mathrm{train}$  \\
\hline
\multirow{8}{*}{118} & \multirow{2}{*}{zeroth-order} & \emph{$M_{\mathrm{FCNN}}$} & \multirow{2}{*}{0.02MB} & 2s & 6min  \\  
& & \emph{${M}_{\mathrm{STRT}}$} & & 2s & 6min  \\
\cline{2-6} 
& Jacobian & \emph{${M}_{\mathrm{DC3}}$} & 14.4MB & 29s & 48min  \\  
\cline{2-6}
& \multirow{5}{*}{batch mean} & $M_0$ & 0.58MB & 17s & 28min  \\ 
\cline{3-4} 
& & $M_1$ & \multirow{4}{*}{0.47MB} & 11s & 18min  \\ 
& & $M_2$ & & 4s & 7min \\ 
& & $M_3$ & & 11s & 18min \\ 
& & $M_4$ & & 4s & 7min \\ 
\hline 

\multirow{8}{*}{1354} & \multirow{2}{*}{zeroth-order} & \emph{$M_{\mathrm{FCNN}}$} & \multirow{2}{*}{0.13MB} & 2.1min & 17.5h  \\  
& & \emph{${M}_{\mathrm{STRT}}$} & & 1.3min & 4.3h\\
\cline{2-6} 
& Jacobian & \emph{${M}_{\mathrm{DC3}}$} & 1.87GB & 16.0min & 6.6h \\  
\cline{2-6}
& \multirow{5}{*}{batch mean} & $M_0$ & 0.06GB & 3.5min & 5.8h \\ 
\cline{3-4} 
& & $M_1$ & \multirow{4}{*}{0.05GB} & 2.4min & 4.0h \\ 
& & $M_2$ & & 1.5min & 2.5h \\ 
& & $M_3$ & & 2.2min & 3.6h\\ 
& & $M_4$ & & 1.3min & 2.1h \\ 
\hline 

\multirow{8}{*}{2869} & \multirow{2}{*}{zeroth-order} & \emph{$M_{\mathrm{FCNN}}$} & \multirow{2}{*}{0.260MB} & 7.9min & 26.3h \\  
& & \emph{${M}_{\mathrm{STRT}}$} & & 6.1min & 10.1h \\
\cline{2-6} 
& Jacobian & \emph{${M}_{\mathrm{DC3}}$} & 8.432GB & 89.0min & 14.8h  \\  
\cline{2-6}
& \multirow{5}{*}{batch mean} & $M_0$ & 0.266GB & 10.9min & 101min \\ 
\cline{3-4} 
& & $M_1$ & \multirow{4}{*}{0.263GB} & 8.1min & 75min\\ 
& & $M_2$ & & 6.0min & 56min \\ 
& & $M_3$ & & 7.0min & 62min\\ 
& & $M_4$ & & 4.9min & 47min \\ 
\hline 

\multirow{8}{*}{9241} & \multirow{2}{*}{zeroth-order} & \emph{$M_{\mathrm{FCNN}}$} & \multirow{2}{*}{0.73MB} & \xmark & \xmark \\  
& & \emph{${M}_{\mathrm{STRT}}$} & & 39.3 min & 10.1h \\
\cline{2-6} 
& Jacobian & \emph{${M}_{\mathrm{DC3}}$} & 87.457GB & \xmark & \xmark \\  
\cline{2-6}
& \multirow{5}{*}{batch mean} & $M_0$ & 2.745GB & 49.5min & 8.2h \\ 
\cline{3-4} 
& & $M_1$ & \multirow{4}{*}{2.737GB} & 44.5min & 7.4h\\ 
& & $M_2$ & & 32.7min & 5.4h  \\ 
& & $M_3$ & & 32.4min & 5.4h \\ 
& & $M_4$ & & 31.1min & 5.1h \\ 
\hline 

\end{tabular}
}
\end{table}

\begin{table}[ht!]
\caption{The comparison results of the total computational time on all testing samples.}
\label{test_time}
\centering
\scalebox{1.1}{
\begin{tabular}{c|c|c}\hline 
System & $T_\mathrm{prop}$ & $T_\mathrm{opt}$ \\
\hline
118 & 0.3s & 3.4min \\  
\hline 
1354 & 10.4s & 37.7min \\  
\hline 
2869 & 50.2s & 109.7min \\   
\hline 
9241 & 361.7s & 209.1min \\  
\hline 
\end{tabular}
}
\end{table}

\subsubsection{Performance comparison results}
Table \ref{Performance_comp} shows the performance of our proposed schemes and competing methods. For PEGASE-1354/-2869/-9241 bus systems, the proposed $M_0$--$M_4$ achieve smaller constraint violations. Note that the training fails for \emph{$M_{\mathrm{FCNN}}$}, \emph{${M}_{\mathrm{DC3}}$} and \emph{${M}_{\mathrm{DUAL}}$} on the PEGASE-9241 system because the follow-up FDPF solver cannot find feasible PF solutions. 
Furthermore, compared with $M_{\mathrm{FCNN}}$ and $M_{\mathrm{DC3}}$, $M_0$--$M_4$ have significantly smaller optimality gaps. The proposed approaches result in solutions that demonstrate improved feasibility compared to $M_{\mathrm{STRT}}$, while only sacrificing little in optimality. Moreover, our schemes effectively address the issue of load mismatch by reconstructing the partial decision variables from the FDPF solver. \emph{${M}_{\mathrm{COMP}}$} and \emph{${M}_{\mathrm{LSTM}}$} have achieved promising results in minimizing $l_{\mathrm{cost}}$ because of the supervised loss of active power generation outputs. Both methods adopt VS3 scheme which requires a PF solver in the post-processing phase to satisfy the power flow balance equations. However, the inequality constraints cannot be satisfied, resulting in significant violations. LSTM architectures excel in capturing long-term dependencies in sequential data, while FCNNs demonstrate robust capabilities for universal function approximation in single-period OPF analysis with non-temporal data. Additionally, the performance of $M_1$ is similar to that of $M_0$, indicating that removing unlikely violated constraints does not significantly impact training accuracy. $M_1$ and $M_2$ have comparable performance, which indicates that the linearized Jacobian approximation is sufficiently accurate for training purposes. Finally, $M_2$ slightly outperforms $M_4$ that uses the decoupled model. 

Table \ref{time} shows the training time and the storage requirements (one batch of training samples) for different gradient computation methods. Our proposed $M_2$ and $M_4$ are capable of completing the training (including hyperparameter tuning) one day in advance. When considering the benchmark systems (excluding the PEGASE-9241 system), the speedup ratios of $M_2$ and $M_4$ compared with $M_{\mathrm{DC3}}$ for the training time $t_\mathrm{train}$ are 7x/7x, 10x/12x, and 14x/18x, respectively. This clearly demonstrates the computational advantages of our proposed linearized (decoupled) Jacobian, particularly as the size of the power grid increases. Furthermore, the per-epoch training times of $M_2$ and $M_4$ are similar to those of $M_{\mathrm{FCNN}}$ and $M_{\mathrm{STRT}}$. The total training time consists of the data preparation time and the neural network training time. Based on Tables \ref{Methods_time} and \ref{time}, our proposed methods can complete the training process one day ahead and achieve daily updates based on the latest data instances. By shifting the computational burden offline, the neural network can rapidly solve the optimal power flow problem for the actual load demand in the real-time market. Considering the varying load demand, a short market clearing time enables the ISO to optimize energy dispatch promptly. As shown in Table \ref{test_time}, the trained neural network can be used as a rapid online solver to complete the market clearing in seconds or minutes.

The proposed techniques demonstrate the ability to achieve optimal solutions with fewer training epochs, resulting in reduced total training times compared to using zeroth-order gradient estimation. Furthermore, when compared to $M_0$, $M_1$ exhibits reduced computational time and memory requirements, validating the effectiveness of the reduced branch set. The computational time of $M_4$ is also lower than that of $M_2$, indicating that the decoupled formulation successfully reduces the computational burden. Ultimately, both $M_2$ and $M_4$ surpass other methods in terms of feasibility and computational efficiency. While $M_2$ outperforms $M_4$ in terms of feasibility, it requires more computational time. Additionally, the swift computational time for testing validates the real-time applicability of our proposed techniques.

\subsubsection{Learning process comparison results}
\begin{figure}[tb!]
    \centering
    \includegraphics[width=0.9\linewidth]{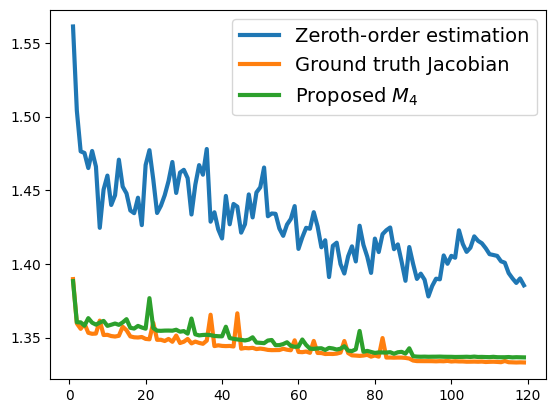}
    \caption{The training loss trajectory of different gradient estimation methods. The learning rate is set to 0.0005 for the initial 90 epochs, then reduced to 0.0001 for the last 30 epochs.}
    \label{lp1}
\end{figure}

Fig. \ref{lp1} shows the training loss trajectory of different gradient computation methods. The gradient computation of the zeroth-order gradient estimation is straightforward. However, it cannot accurately approximate the gradient of non-linear and non-convex power flow and branch flow equations. With the inaccurate gradient descent direction, the zeroth-order estimation method performs worst in the learning process. Compared to the ground truth Jacobian gradient, the proposed gradient estimation method $M_4$ achieves a comparable convergence performance. The proposed physics-informed gradient methods require less training time and storage than the ground truth model, offering advantages when computing resources is limited.

\begin{figure}[tb!]
    \centering
    \includegraphics[width=0.9\linewidth]{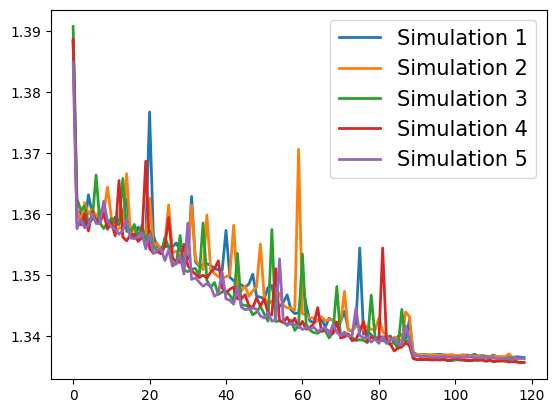}
    \caption{The total training loss trajectory of the proposed method $M_4$ over epochs on the IEEE-118 bus system. The learning rate is 0.0005 in the first 90 epochs and 0.0001 in the last 30 epochs.}
    \label{M4_118}
\end{figure}

\begin{figure}[tb!]
    \centering
    \includegraphics[width=0.9\linewidth]{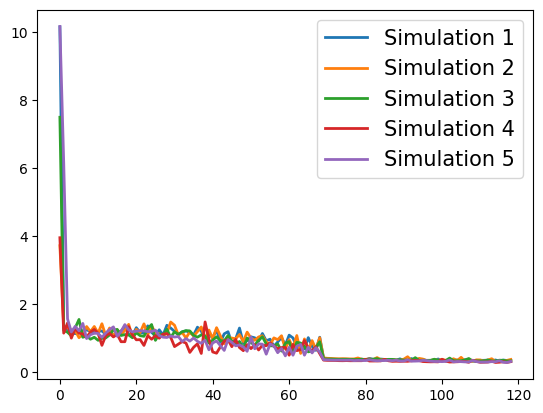}
    \caption{The total training loss evolution process of the proposed method $M_4$ over epochs on the PEGASE-1354 bus system. The learning rate is 0.0005 in the first 70 epochs and 0.0001 in the last 50 epochs.}
    \label{M4_1354}
\end{figure}

We independently train the neural network five times using the proposed method $M_4$. Fig. \ref{M4_118} and Fig. \ref{M4_1354} show the training loss trajectory on the IEEE-30 and PEGASE-1354 bus systems. We observe that the training loss decreases, stabilizes, and eventually converges to a local minimum as the number of training epochs increases. Due to the random initialization of neural network weights, these independent runs converge to different local minima. However, the final training loss values are desirable and closely aligned. The neural network is considered converged when the training loss ceases to decrease significantly over successive epochs and stabilizes within an acceptable range. For instance, in the case of the IEEE-118 bus system, from epoch 100 to epoch 120, the training loss values do not continue to decrease but instead fluctuate within narrow ranges: 0.0005, 0.0008, 0.0008, 0.0007, and 0.0003 for the five runs, respectively. Concurrently, the trained neural network demonstrates promising results on the validation dataset based on optimality and feasibility evaluation metrics. Consequently, we terminate the training at epoch 100 to optimize training time. It is noteworthy that the observed oscillations are attributed to a larger learning rate rather than the proposed gradient estimation technique. With a smaller learning rate, the training loss stabilizes and converges to a local minimum.

\begin{figure}[tb!]
    \centering
    \includegraphics[width=0.95\linewidth]{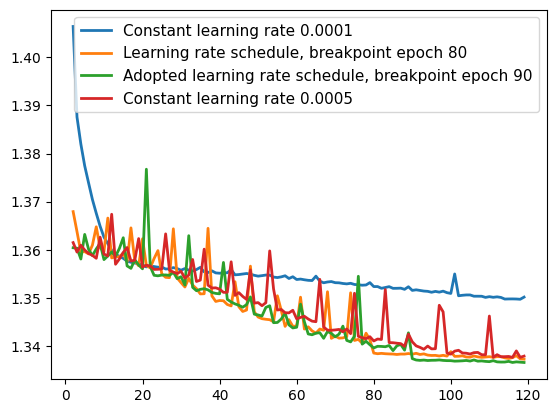}
    \caption{The total training loss trajectory of the proposed method $M_4$ over epochs using different learning rate schedules on the IEEE-118 bus system. The plot starts from the second epoch for better demonstration.}
    \label{learningrate}
\end{figure}

We have tried different learning rates for the learning rate schedule during the hyper-parameter tuning process. As shown in Fig. \ref{learningrate}, a constant learning rate $0.0001$ leads to a slow learning process, while a significant learning rate $0.0005$ has more oscillating. Besides, a considerable breakpoint epoch helps improve training efficiency by using a significant learning rate in more beginning training epochs. We use epoch 90 as the breakpoint instead of epoch 80 because the former performs better in convergence.

\section{Conclusion and Discussion}\label{sec:conclusion}
In this paper, we propose a deep learning framework as a rapid solution to the real-time AC-OPF problem. The framework involves utilizing ridge regression in pseudo-labeling. This technique utilizes the creation of a hybrid dataset, thus reducing the time needed to prepare many ground truth data pairs required by traditional solvers. Additionally, we introduce innovative methods for computing the gradient of reconstructed variables in relation to the neural network outputs. To balance computational complexity and ensure satisfactory training performance, we employ a linearized decoupled Jacobian formulation. Moreover, a batch-mean estimation mechanism is devised specifically for mini-batch training, effectively reducing the computational load associated with calculating the Jacobian tensor. Furthermore, we present a reduced branch set that mitigates the computational complexity arising from the branch flow gradient. \blue{The simulation results validate the effectiveness of the proposed techniques, showing high feasibility and a minimal optimality gap in medium and large-scale power systems. Specifically, the optimality gap is under $1\%$, and average constraint violations are on the order of  $10^{-4}$. For the PEGASE-9241 bus system, the computation time per data instance is under one second, and the proposed learning framework achieves a 35x speed-up over conventional optimization solvers. Additionally, the gradient estimation method accelerates neural network training, providing an average 12x speed-up compared to the ground Jacobian computation.} 

\blue{
The high penetration of renewable energy sources (RESs) in the power grid poses significant challenges to independent system operators. Battery energy storage systems (BESS) enable flexible energy storage and release, essential for integrating RESs. However, incorporating BESS introduces a multi-period OPF problem with dynamic battery constraints, and conventional solvers often lack the speed needed for real-time energy markets and system operations.}

\blue{
To address the uncertainty of renewable generation and enable rapid solutions, we propose a learning-based controller. Future work will explore advanced techniques such as transformers, graph neural networks, diffusion models, and reinforcement learning (RL), which show promise in handling stochastic, multi-period optimization problems. Additionally, large language models (LLMs) may serve as meta-learners or assist in optimization framework design and scenario analysis. These cutting-edge approaches aim to advance multi-period OPF solutions for high-RES penetration scenarios.
} 

\bibliographystyle{IEEEtran}
\bibliography{OPFrefs, IEEEabrv}

\begin{thebibliography}{10}
\providecommand{\url}[1]{#1}
\csname url@samestyle\endcsname
\providecommand{\newblock}{\relax}
\providecommand{\bibinfo}[2]{#2}
\providecommand{\BIBentrySTDinterwordspacing}{\spaceskip=0pt\relax}
\providecommand{\BIBentryALTinterwordstretchfactor}{4}
\providecommand{\BIBentryALTinterwordspacing}{\spaceskip=\fontdimen2\font plus
\BIBentryALTinterwordstretchfactor\fontdimen3\font minus \fontdimen4\font\relax}
\providecommand{\BIBforeignlanguage}[2]{{%
\expandafter\ifx\csname l@#1\endcsname\relax
\typeout{** WARNING: IEEEtran.bst: No hyphenation pattern has been}%
\typeout{** loaded for the language `#1'. Using the pattern for}%
\typeout{** the default language instead.}%
\else
\language=\csname l@#1\endcsname
\fi
#2}}
\providecommand{\BIBdecl}{\relax}
\BIBdecl

\bibitem{Mohammadi}
A.~Mohammadi and A.~Kargarian, ``Accelerated and robust analytical target cascading for distributed optimal power flow,'' \emph{IEEE Trans. Industr. Inform.}, vol.~16, no.~12, pp. 7521--7531, Dec 2020.

\bibitem{Hai-Tao}
H.-T. Zhang, W.~Sun, Y.~Li, D.~Fu, and Y.~Yuan, ``A fast optimal power flow algorithm using powerball method,'' \emph{IEEE Trans. Industr. Inform.}, vol.~16, no.~11, pp. 6993--7003, Nov 2020.

\bibitem{Chatzos2022}
M.~Chatzos, T.~W.~K. Mak, and P.~V. Hentenryck, ``Spatial network decomposition for fast and scalable {AC-OPF} learning,'' \emph{IEEE Trans. Power Syst.}, vol.~37, no.~4, pp. 2601--2612, 2022.

\bibitem{Fioretto2019}
F.~Fioretto, T.~Mak, and P.~Van~Hentenryck, ``Predicting {AC} optimal power flows: Combining deep learning and lagrangian dual methods,'' \emph{Proceedings of the AAAI Conference on Artificial Intelligence}, vol.~34, pp. 630--637, 04 2020.

\bibitem{chatzos2020}
\BIBentryALTinterwordspacing
M.~Chatzos, F.~Fioretto, T.~W.~K. Mak, and P.~V. Hentenryck, ``High-fidelity machine learning approximations of large-scale optimal power flow,'' 2020. [Online]. Available: \url{https://arxiv.org/abs/2006.16356}
\BIBentrySTDinterwordspacing

\bibitem{Seonho2023}
S.~Park, W.~Chen, T.~W. Mak, and P.~Van~Hentenryck, ``Compact optimization learning for ac optimal power flow,'' \emph{IEEE Trans. Power Syst.}, vol.~39, no.~2, pp. 4350--4359, 2024.

\bibitem{Huang2021_v}
W.~Huang, X.~Pan, M.~Chen, and S.~H. Low, ``{DeepOPF-V}: Solving {AC-OPF} problems efficiently,'' \emph{IEEE Trans. Power Syst.}, vol.~37, pp. 800--803, 2021.

\bibitem{Lei2021}
X.~Lei, Z.~Yang, J.~Yu, J.~Zhao, Q.~Gao, and H.~Yu, ``Data-driven optimal power flow: A physics-informed machine learning approach,'' \emph{IEEE Trans. Power Syst.}, vol.~36, no.~1, pp. 346--354, 2021.

\bibitem{Owerko2020}
D.~Owerko, F.~Gama, and A.~Ribeiro, ``Optimal power flow using graph neural networks,'' in \emph{IEEE ICASSP}, 2020, pp. 5930--5934.

\bibitem{falconer2022}
T.~Falconer and L.~Mones, ``Leveraging power grid topology in machine learning assisted optimal power flow,'' \emph{IEEE Trans. Power Syst.}, vol.~38, no.~3, pp. 2234--2246, 2023.

\bibitem{Gao2023}
M.~Gao, J.~Yu, Z.~Yang, and J.~Zhao, ``A physics-guided graph convolution neural network for optimal power flow,'' \emph{IEEE Trans. Power Syst.}, pp. 1--11, 2023.

\bibitem{NELLIKKATH}
R.~Nellikkath and S.~Chatzivasileiadis, ``Physics-informed neural networks for {AC} optimal power flow,'' \emph{Electric Power Systems Research}, vol. 212, p. 108412, 2022.

\bibitem{deepopf}
X.~Pan, M.~Chen, T.~Zhao, and S.~H. Low, ``{DeepOPF}: A feasibility-optimized deep neural network approach for {AC} optimal power flow problems,'' \emph{IEEE Systems Journal}, vol.~17, no.~1, pp. 673--683, 2023.

\bibitem{WANG2022}
Z.~Wang, J.-H. Menke, F.~Schäfer, M.~Braun, and A.~Scheidler, ``Approximating multi-purpose {AC} optimal power flow with reinforcement trained artificial neural network,'' \emph{Energy and AI}, vol.~7, p. 100133, 2022.

\bibitem{Zhou2023}
M.~Zhou, M.~Chen, and S.~H. Low, ``{DeepOPF-FT}: One deep neural network for multiple {AC-OPF} problems with flexible topology,'' \emph{IEEE Trans. Power Syst.}, vol.~38, no.~1, pp. 964--967, 2023.

\bibitem{ZhangIot}
Z.~Zhang, R.~Deng, D.~K.~Y. Yau, and P.~Chen, ``Zero-parameter-information data integrity attacks and countermeasures in iot-based smart grid,'' \emph{IEEE Internet Things J.}, vol.~8, no.~8, pp. 6608--6623, 2021.

\bibitem{huang2021}
W.~Huang and M.~Chen, ``{DeepOPF-NGT}: A fast unsupervised learning approach for solving {AC-OPF} problems without ground truth,'' in \emph{ICML Workshop on Tackling Climate Change with Machine Learning}, 2021.

\bibitem{Junfei2022}
J.~Wang and P.~Srikantha, ``Fast optimal power flow with guarantees via an unsupervised generative model,'' \emph{IEEE Trans. Power Syst.}, pp. 1--12, 2022.

\bibitem{parkaaai}
S.~Park and P.~Van~Hentenryck, ``Self-supervised primal-dual learning for constrained optimization,'' in \emph{Proceedings of the AAAI Conference on Artificial Intelligence}, vol.~37, no.~4, 2023, pp. 4052--4060.

\bibitem{Cao}
D.~Cao, W.~Hu, X.~Xu, Q.~Wu, Q.~Huang, Z.~Chen, and F.~Blaabjerg, ``Deep reinforcement learning based approach for optimal power flow of distribution networks embedded with renewable energy and storage devices,'' \emph{J. Mod. Power Syst. Clean Energy}, vol.~9, no.~5, pp. 1101--1110, 2021.

\bibitem{donti2021dc3}
P.~Donti, D.~Rolnick, and J.~Z. Kolter, ``{DC3}: A learning method for optimization with hard constraints,'' in \emph{ICML}, 2021.

\bibitem{KejunGlobal}
K.~Chen, S.~Bose, and Y.~Zhang, ``Unsupervised deep learning for {AC} optimal power flow via lagrangian duality,'' in \emph{IEEE Global Communications Conference}, 2022, pp. 5305--5310.

\bibitem{Liu2020}
S.~Liu, P.-Y. Chen, B.~Kailkhura, G.~Zhang, A.~O. Hero~III, and P.~K. Varshney, ``A primer on zeroth-order optimization in signal processing and machine learning: Principals, recent advances, and applications,'' \emph{IEEE Signal Process. Mag.}, vol.~37, no.~5, pp. 43--54, 2020.

\bibitem{yangf2023}
F.~Yang, Z.~Ling, Y.~Zhang, X.~He, Q.~Ai, and R.~C. Qiu, ``Event detection, localization, and classification based on semi-supervised learning in power grids,'' \emph{IEEE Trans. Power Syst.}, vol.~38, no.~5, pp. 4080--4094, 2023.

\bibitem{Farajzadeh}
M.~Farajzadeh-Zanjani, E.~Hallaji, R.~Razavi-Far, M.~Saif, and M.~Parvania, ``Adversarial semi-supervised learning for diagnosing faults and attacks in power grids,'' \emph{IEEE Trans. on Smart Grid}, vol.~12, no.~4, pp. 3468--3478, 2021.

\bibitem{deka2019}
D.~Deka and S.~Misra, ``Learning for {DC-OPF}: Classifying active sets using neural nets,'' in \emph{2019 IEEE Milan PowerTech}, 2019, pp. 1--6.

\bibitem{robson2020}
\BIBentryALTinterwordspacing
A.~Robson, M.~Jamei, C.~Ududec, and L.~Mones, ``Learning an optimally reduced formulation of {OPF} through meta-optimization,'' \emph{arXiv preprint arXiv:1911.06784}, 2019. [Online]. Available: \url{https://arxiv.org/abs/1911.06784}
\BIBentrySTDinterwordspacing

\bibitem{Yang2017}
J.~Yang, N.~Zhang, C.~Kang, and Q.~Xia, ``A state-independent linear power flow model with accurate estimation of voltage magnitude,'' \emph{IEEE Trans. Power Syst.}, vol.~32, no.~5, pp. 3607--3617, 2017.

\bibitem{Li2022}
M.~Li, Y.~Du, J.~Mohammadi, C.~Crozier, K.~Baker, and S.~Kar, ``Numerical comparisons of linear power flow approximations: Optimality, feasibility, and computation time,'' in \emph{IEEE PES GM}, 2022, pp. 1--5.

\bibitem{Prechelt2012}
L.~Prechelt, \emph{Early Stopping-But When?}\hskip 1em plus 0.5em minus 0.4em\relax Berlin, Heidelberg: Springer, 2012.

\end{thebibliography}



\end{document}